\documentclass[10pt,letterpaper]{article}
\usepackage[top=0.85in,left=2.75in,footskip=0.75in]{geometry}

\usepackage{amsmath,amssymb}

\usepackage{changepage}

\usepackage[utf8x]{inputenc}

\usepackage{textcomp,marvosym}

\usepackage{cite}

\usepackage{nameref,hyperref}
\usepackage{graphicx}
\usepackage[labelformat=empty, position=top]{subcaption}
\usepackage{bm}
\usepackage[export]{adjustbox}

\usepackage{blindtext}
\usepackage{comment}

\usepackage[right]{lineno}

\usepackage{microtype}
\DisableLigatures[f]{encoding = *, family = * }

\usepackage[table]{xcolor}

\usepackage{array}

\newcolumntype{+}{!{\vrule width 2pt}}

\newlength\savedwidth

\usepackage[capitalise]{cleveref}

\usepackage[inline]{enumitem}
\usepackage{comment}

\usepackage[labelformat=empty, position=top]{subcaption}

\raggedright
\usepackage[aboveskip=1pt,labelfont=bf,labelsep=period,justification=raggedright,singlelinecheck=off]{caption}

\makeatletter
\renewcommand{\@biblabel}[1]{\quad#1.}
\makeatother

\usepackage{lastpage,fancyhdr,graphicx}
\usepackage{epstopdf}
\fancyheadoffset[L]{2.25in}
\fancyfootoffset[L]{2.25in}
\begin{document}
	\vspace*{0.2in}
	
	\begin{flushleft}
		{\Large
			\textbf\newline{STDP and the distribution of preferred phases in the whisker system} 
		}
		\newline
		\\
		Nimrod Sherf\textsuperscript{1,2*},
		Maoz Shamir\textsuperscript{1,2,3}
		\\
		\bigskip
		\textbf{1}	Physics Department, Ben-Gurion University of the Negev, Beer-Sheva, Israel
		\\
		\textbf{2} Zlotowski Center for Neuroscience, Ben-Gurion University of the Negev, Beer-Sheva, Israel
		\\
		\textbf{3} Department of Physiology and Cell Biology Faculty of Health Sciences, Ben-Gurion University of the
		Negev, Beer-Sheva, Israel
		\\
		\bigskip
		
		%
		%
		
		
		
		
		
		* sherfnim@post.bgu.ac.il
		
	\end{flushleft}
	\section*{Abstract}
Rats and mice use their whiskers to probe the environment. By rhythmically swiping their whiskers back and forth they can detect the existence of an object, locate it, and identify its texture. Localization can be accomplished by inferring the whisker’s position. Rhythmic neurons that track the phase of the whisking cycle encode information about the azimuthal location of the whisker. These neurons are characterized by preferred phases of firing that are narrowly distributed. Consequently, pooling the rhythmic signal from several upstream neurons is expected to result in a much narrower distribution of preferred phases in the downstream population, which however has not been observed empirically.
Here, we show how spike timing dependent plasticity (STDP) can provide a solution to this conundrum. We investigated the effect of STDP on the utility of a neural population to transmit rhythmic information downstream using the framework of a modeling study. We found that under a wide range of parameters, STDP facilitated the transfer of rhythmic information despite the fact that all the synaptic weights remained dynamic. As a result, the preferred phase of the downstream neuron was not fixed, but rather drifted in time at a drift velocity that depended on the preferred phase, thus inducing a distribution of preferred phases. We further analyzed how the STDP rule governs the distribution of preferred phases in the downstream population. This link between the STDP rule and the distribution of preferred phases constitutes a natural test for our theory.

	\section*{Author summary}
The distribution of preferred phases of whisking neurons in the somatosensory system of rats and mice presents a conundrum: a simple pooling model predicts a distribution that is an order of magnitude narrower than what is observed empirically. Here, we suggest that this non-trivial distribution may result from activity-dependent plasticity in the form of spike timing dependent plasticity (STDP). We show that under STDP, the synaptic weights do not converge to a fixed value, but rather remain dynamic. As a result, the preferred phases of the whisking neurons vary in time, hence inducing a non-trivial distribution of preferred phases, which is governed by the STDP rule. Our results imply that the considerable synaptic volatility which has long been viewed as a difficulty that needs to be overcome, may actually be an underlying principle of the organization of the central nervous system.

	
	\section*{Introduction}
The whisker system is used by rats and mice to actively gather information about their proximal  environment \cite{isett2020cortical,kleinfeld2011neuronal,ahissar2008object,diamond2008and}.
Information about whisker position, touch events, and texture is relayed downstream the somatosensory system via several tracks; in particular, the lemniscal pathway that relays information about both whisking and touch \cite{severson2017active,moore2015vibrissa,campagner2016prediction,urbain2015whisking,wallach2016going,szwed2003encoding,yu2006parallel}.

	\begin{figure} \hypertarget{fig 1}{}
		\begin{adjustwidth}{-6.5cm}{-1cm}
		\begin{subfigure}[t]{0.007\textwidth}\vspace{0.02\textwidth}
		\textbf{(a)} 
	\end{subfigure}
	\begin{subfigure}[t]{0.4\textwidth}  \vspace{0.8cm}\hspace{-0.6cm}
		\includegraphics[width=\linewidth, valign=t]{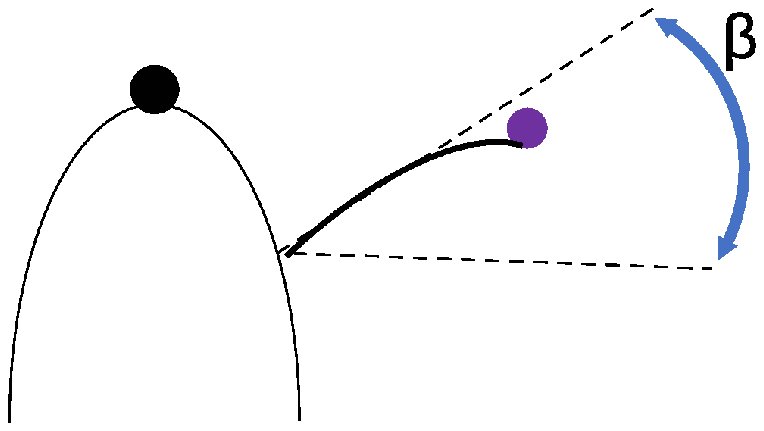}\\[3pt]
		\caption{}
		\label{fig:Fig1a}
	\end{subfigure}\hspace{0.05cm} \vspace{0.05\textwidth}
	\begin{subfigure}[t]{0.4\textwidth}  \vspace{-0.1cm} \hspace{-0.65cm}
		\includegraphics[width=1.5\linewidth, valign=t]{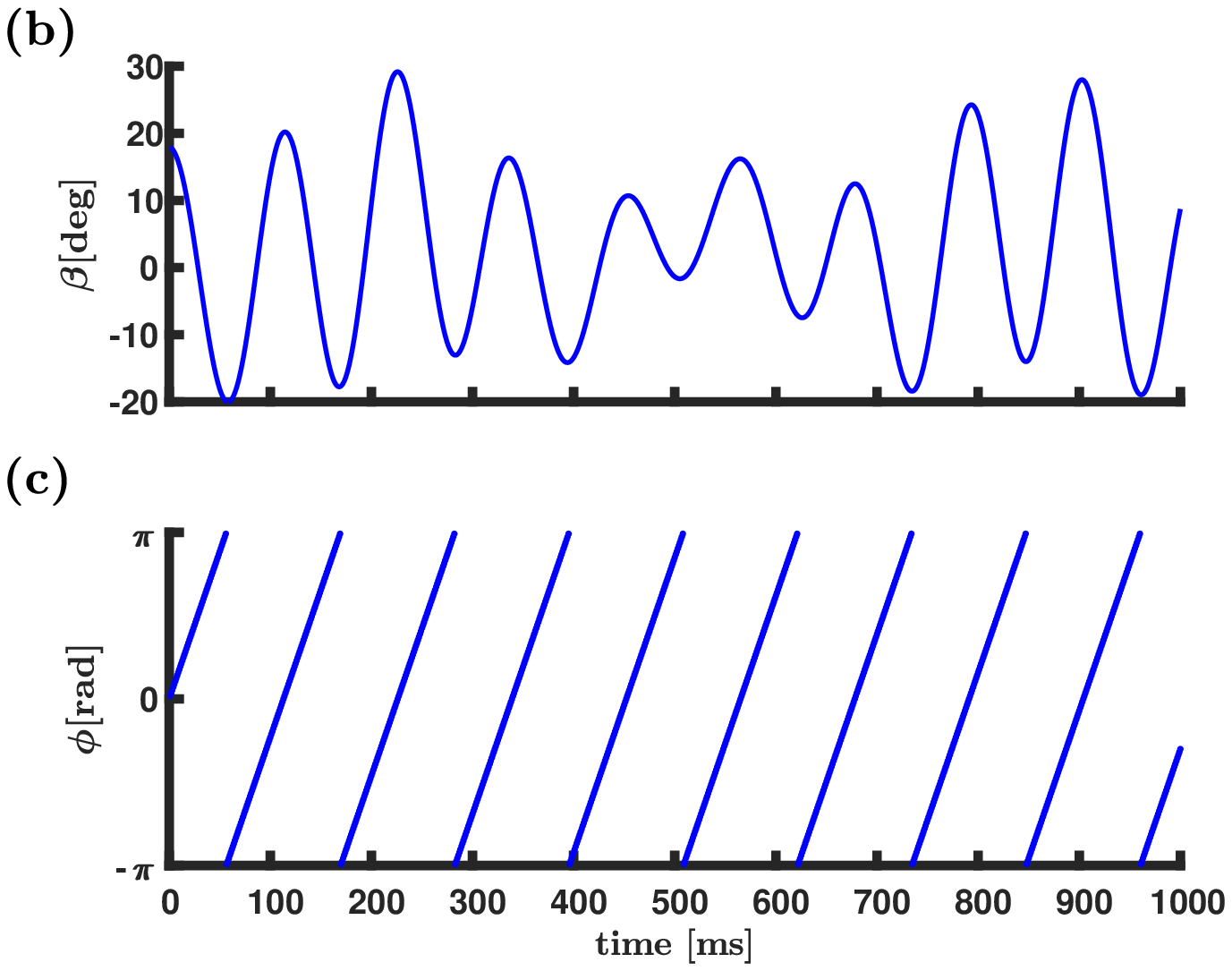}\\[3pt] 
		\caption{}
		\label{fig:Fig1bc}
	\end{subfigure}\hspace{0.3cm} \vspace{0.002\textwidth}
	\hfill \vspace{0.0001\textwidth}
				\begin{subfigure}[t]{0.4\textwidth}  \vspace{-0.1cm}  
					\includegraphics[width=1.5\linewidth, valign=t]{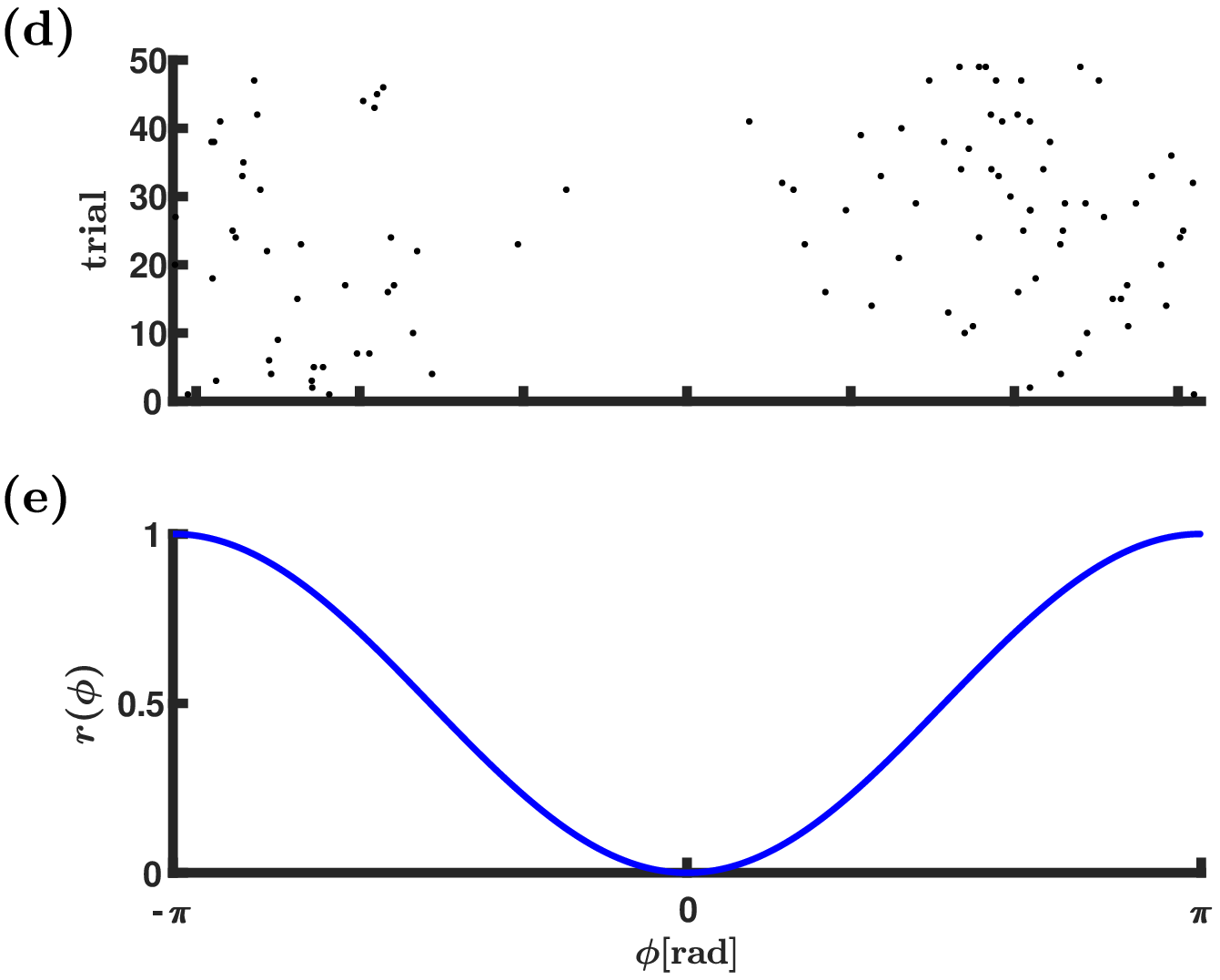}\\[3pt]
					\caption{}
					\label{fig:Fig1de}
				\end{subfigure}\hfill \vspace{0.0001\textwidth}
	\hfill \vspace{0.0001\textwidth}
				
		\caption{ {\bf Representation of the whisking phase}. (a) Mice and rats can infer the azimuthal location of an object by touch.  (b) The angular position of a whisker, $\beta$, (during whisking) is shown as a function of time. The angle is often modeled as $\beta(t)=\beta_{\text{midpoint}}(t)+\Delta \beta(t)\cos(\phi(t))$, where $\beta_{\text{midpoint}}(t)$ and $\Delta \beta(t)$ are the midpoint and the whisking amplitude, respectively. (c) The whisking phase $\phi$ as a function of time is $\phi(t)=(\nu t)_{\text{mod} 2\pi}$, where $\nu$ is the angular frequency of the whisking. (d) \& (e) Raster plot and normalized tuning curve of a neuron with a preferred phase near maximal retraction.}
				\label{fig:intro}
						\end{adjustwidth}
\end{figure}

\begin{figure}
		\begin{subfigure}[t]{0.007\textwidth}
		\textbf{(a)} 
	\end{subfigure}
	\begin{subfigure}[t]{0.45\textwidth}  \vspace{0.05\textwidth}
		\includegraphics[width=\linewidth, valign=t]{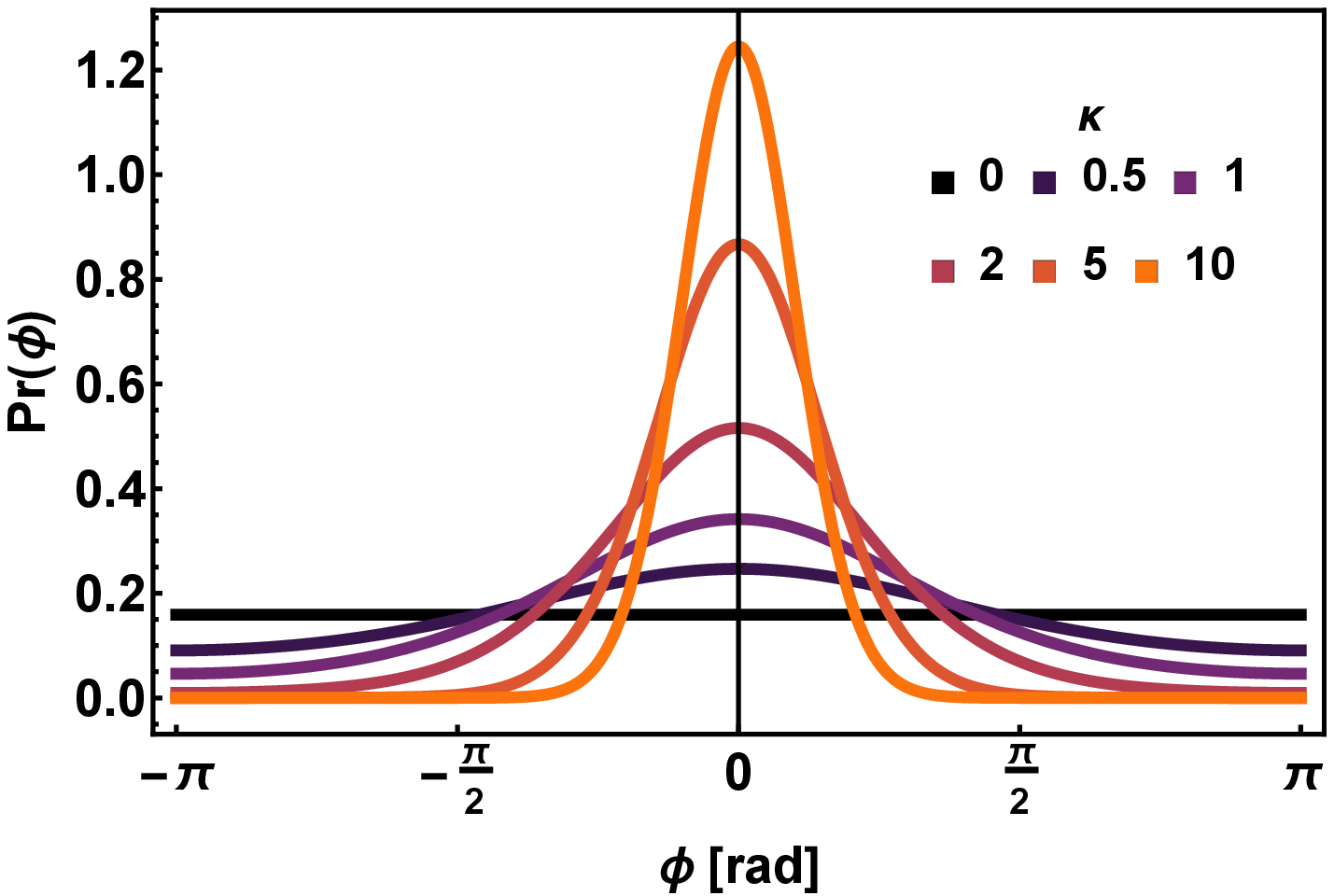}\\[3pt]
		\caption{}
		\label{fig:Fig2a}
	\end{subfigure}\hspace{0.5cm} \vspace{0.02\textwidth}
	\begin{subfigure}[t]{0.007\textwidth}
		\textbf{(b)} 
	\end{subfigure}
	\begin{subfigure}[t]{0.42\textwidth}  \vspace{0.007\textwidth}
		\includegraphics[width=\linewidth, valign=t]{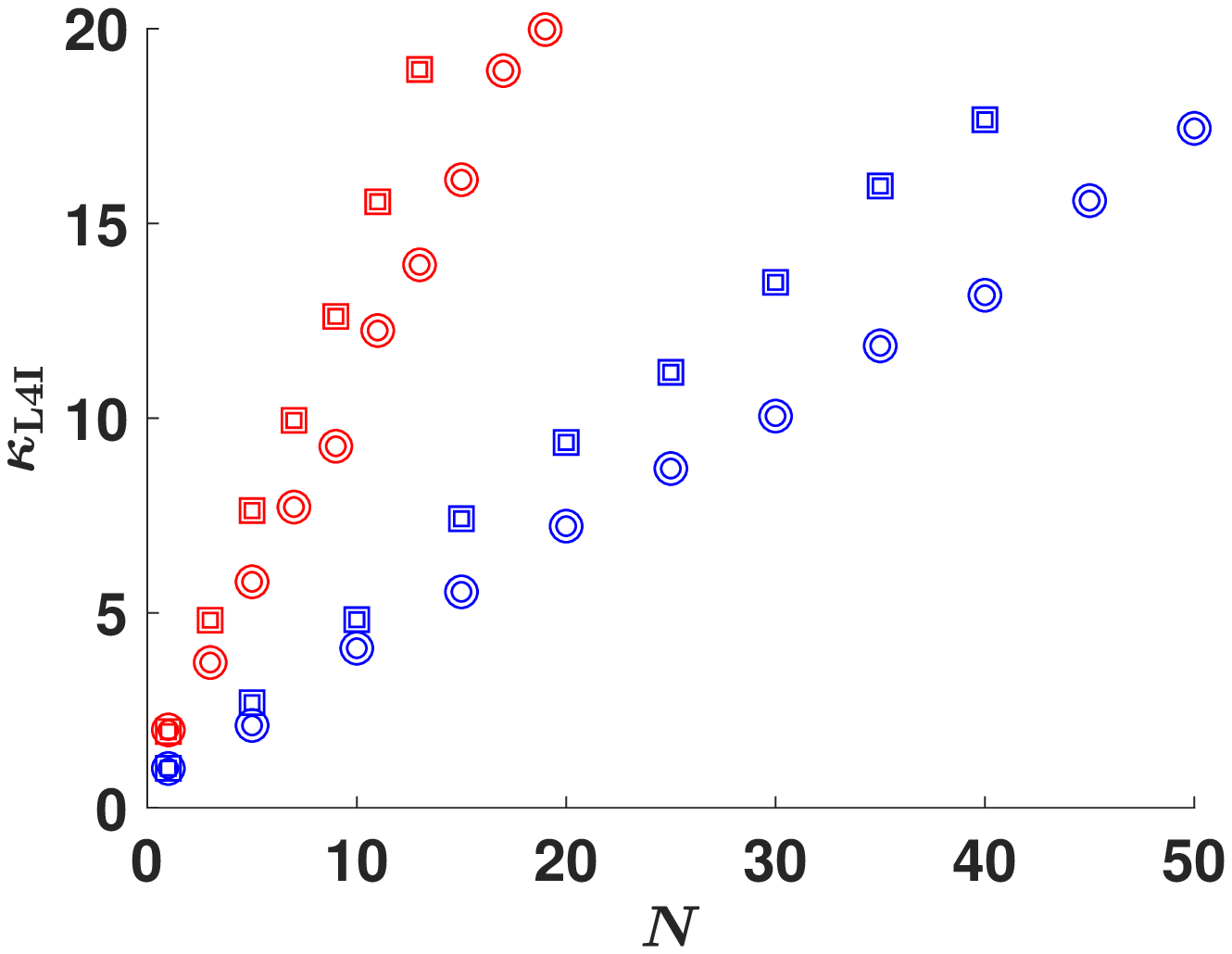}\\[3pt]
		\caption{}
		\label{fig:Fig2b}
	\end{subfigure}\hspace{0.5cm} \vspace{0.02\textwidth}
	\hfill \vspace{0.0001\textwidth}
		\caption{ {\bf Distribution of preferred phases}. (a) The von-Mises distribution, \cref{eq:VM}, with $\psi = 0$ is shown for different values of $\kappa$ as depicted by color. (b) Distribution width in the uniform/random pooling model. The width of the distribution of preferred phases, $\kappa_{ \mathrm{L4I}}$, in the downstream layer (L4I) is shown as a function of the number of pooled VPM neurons, $N$, for the uniform and random pooling in squares and circles, respectively. The width of the distribution, $\kappa_{ \mathrm{L4I}}$, was estimated from 10000 repetitions of drawing $N$ preferred phases of the upstream population with $\kappa = 1$ (blue) and $\kappa = 2$ (red). }
\end{figure}

During whisking, the animal moves its vibrissae back and forth in a rhythmic manner  \hyperlink{fig 1}{Fig. 1a-1c}. Neurons that track the azimuthal position of the whisker by firing in a preferential manner to the phase of the whisking cycle are termed whisking neurons. Whisking neurons in the ventral posteromedial nucleus (VPM) of the thalamus as well as inhibitory whisking neurons in layer 4 of the barrel cortex are characterized by a preferred phase at which they fire with the highest rate during the whisking cycle \cite{ego2012coding,moore2015vibrissa,yu2016layer,grion2016coherence,yu2019recruitment,gutnisky2017mechanisms,isett2020cortical} \hyperlink{fig 1}{Fig. 1d-1e}. The distribution of preferred phases is non-uniform and can be approximated by the circular normal (Von-Mises) distribution
\begin{equation}\label{eq:VM}
    \Pr (\phi) = \frac{e^{\kappa \cos(\phi-\psi)}}{2 \pi I_{0}(\kappa)}
\end{equation}
where $\psi$ is the mean phase and $I_0(\kappa)$ is the modified Bessel function of order 0, \cref{fig:Fig2a}. The parameter $\kappa$ quantifies the width of the distribution; $\kappa = 0$ yields a uniform (flat) distribution, whereas in the limit of $\kappa \rightarrow \infty$ the distribution converges to a delta function. Typical values for $\kappa$ in the thalamus and for layer 4 inhibitory whisking neurons are $\kappa_{ \mathrm{VPM}} \approx \kappa_{ \mathrm{L4I}} \approx 1$ where $\psi_{ \mathrm{VPM}} \approx 5 \pi /6$ [rad] and  $\psi_{ \mathrm{L4I}} \approx 0.5$ [rad] \cite{lefort2009excitatory}. 

Assuming the rhythmic input to L4I neurons originates solely from the VPM, the distribution width of preferred phases of L4I neurons can be computed. \cref{fig:Fig2b} shows the expected distribution width, $\kappa_{ \mathrm{L4I}}$, as a function of the number of VPM neurons that serve as input to single L4I neurons, for uniform (squares) and random (circles) pooling. We shall term this naive pooling the `pooling model' hereafter. As can be seen from the figure, even a random pooling of $N=10$ results in a distribution of preferred phases that is considerably narrower than empirically observed. This result is particularly surprising since the number of thalamic neurons synapsing onto a single L4 neuron was estimated to be on the order of 100, see e.g.\ \cite{yu2016layer,gutnisky2017mechanisms}.

Recently, the effects of rhythmic activity on the spike timing dependent plasticity (STDP) dynamics of feed-forward synaptic connections have been examined \cite{luz2016oscillations,sherf2020multiplexing}. It was shown that in this case the synaptic weights remain dynamic. As a result, the phase of the downstream neuron is arbitrary and drifts in time; thus, effectively, inducing a distribution of preferred phases in the downstream population. However, if the phases of the downstream population are arbitrary and drift in time, how can information about the whisking phase be transmitted?

Here we investigated the hypothesis that the distribution of phases in a downstream layer is governed by the interplay of the distribution in the upstream layer and the STDP rule. The remainder of this article is organized as follows. 
First, we define the network architecture and the STDP learning rule. We then derive a mean-field approximation for the STDP dynamics in the limit of a slow learning rate for a threshold-linear Poisson downstream neuron model. 
Next, we show that STDP dynamics can generate non-trivial distributions of preferred phases in the downstream population and analyze how the parameters characterizing the STDP govern this distribution. Finally, we summarize the results, discuss the limitations of this study and suggest essential empirical predictions of our theory.

	\section*{Results}  \label{Results}
	

	\subsection*{The upstream thalamic population}
	
	We model a system of $N$ thalamic  excitatory whisking neurons, synapsing in a feed-forward manner onto a single inhibitory L4 barrel cortical neuron. Unless stated otherwise, following \cite{gutnisky2017mechanisms}, in our numerical simulations we used $N=150$.
	The spiking activity of the thalamic neurons is modelled by independent inhomogeneous Poisson processes  with an instantaneous firing rate that follows the whisking cycle:
	\begin{equation}\label{eq:meanfiring}
\langle \rho_{k}(t) \rangle=D(1+\gamma \cos[ \nu t-\phi_{k}]),
\end{equation} 
where $ \rho_{k}(t)=\sum_{i} \delta(t-t_{k,i})$, $k \in \left\lbrace 1, ... N \right\rbrace $, is the spike train of the $k$th thalamic neuron, with  $\lbrace t_{k,i} \rbrace _{i=1}^{\infty}$ denoting its spike times.
The parameter $D$ is the mean firing rate during whisking (averaged over one cycle), $\gamma$ is the modulation depth, $\nu$ is the angular frequency of the whisking, and $\phi_{k}$ is the preferred phase of firing of the $k$th thalamic neuron. We further assume that the preferred phases in the thalamic population, $ \{ \phi_k \}_{k=1}^N$, are distributed  i.i.d.\ according to the von-Mises distribution, \cref{eq:VM}.

	\subsection*{The downstream layer 4 inhibitory neuron model}  \label{subsec:neuroDyn}

	To facilitate the analysis we model the response of the downstream layer 4 inhibitory (L4I) neuron to its thalamic inputs by the linear Poisson model, which has been frequently used in the past \cite{softky1993highly,shadlen1998variable,masquelier2009oscillations,luz2016oscillations,morrison2008,song2000competitive,kempter2001intrinsic,kempter1999hebbian,luz2012balancing}. 
	Given the thalamic responses, the firing of the L4I neuron follows inhomogeneous Poisson process statistics with instantaneous firing rate 
	\begin{equation}\label{eq:meanpost}
	r_{ \mathrm{L4I}} (t)=\frac{1}{N} \sum_{k=1}^{N}w_{k}  \rho_{k}(t-d),
	\end{equation}
	where $ d>0$ represents a characteristic delay, and $w_{k}$ is the synaptic weight of the $k$th  thalamic neuron.
	 
Due to the rhythmic nature of the thalamic inputs, \cref{eq:meanfiring}, and the linearity of the L4I neuron, \cref{eq:meanpost}, the L4I neuron will exhibit rhythmic activity:
\begin{equation}
\label{eq:L4IMeanFiring}
    \langle r_{ \mathrm{L4I}}(t) \rangle = D_{ \mathrm{L4I}}(1+\gamma_{ \mathrm{L4I}} \cos[ \nu t-\psi_{ \mathrm{L4I}}]),
\end{equation} 
with a mean, $D_{\text{L4I}}$, a modulation depth,  $\gamma_{\text{L4I}}$, and a preferred phase  $\psi_{\text{L4I}}$, that depend on global order parameters that characterize the thalamocortical synaptic weights population. For large $N$ these order parameters are given by:
	\begin{equation}\label{eq:wbar}
	\bar{w}(t)= \int_{0}^{2 \pi} \Pr(\phi) w(\phi, t)  d \phi
	\end{equation}
	and
	\begin{equation}\label{eq:wtilda}
	\tilde{w}(t) e^{i \psi}= \int_{0}^{2 \pi}  \Pr(\phi) w(\phi, t)e^{i\phi}d \phi.
	\end{equation}
where  $\bar{w}$ is  the mean synaptic weight and  $\tilde{w}e^{i \psi}$ is its first Fourier component. The phase $\psi$  is determined by the condition that $\tilde{w}$ is real and non-negative.
Consequently, the L4I neurons in our model respond to whisking with a mean $D_{\text{L4I}} = D \bar{w}$, a modulation depth of $\gamma_{\text{L4I}} = \gamma \tilde{w} / \bar{w}$, and a preferred phase  $\psi_{\text{L4I}}=\psi+\nu d$.	
	
	\subsection*{The STDP rule}  \label{subsec:STDP rule}
We model the modification of the synaptic weight,  $\Delta w$,  following either a pre- or post-synaptic spike as a sum of two processes: potentiation (+) and depression (-)  \cite{luz2014effect,gutig2003learning,luz2012balancing}, as 
	\begin{equation}\label{eq:deltaw}
	\Delta{w}=\lambda[f_{+}(w)K_{+}(\Delta t)-f_{-}(w)K_{-}(\Delta t)].
	\end{equation}
The parameter $\lambda$ denotes the learning rate. We further assume separation of variables and write each term (potentiation and depression) as the product of the function of the synaptic weight, $f_{\pm}(w)$, and the temporal kernel of the STDP rule, $K_{\pm}(\Delta t)$, where $\Delta t = t_\text{post}-t_\text{pre}$ is the time difference between pre- and post-synaptic spike times. Following G\"utig et al.  \cite{gutig2003learning} the weight dependence functions, $f_{\pm}(w)$, were chosen to be:
 	\begin{subequations}\label{eq:fplusminus}
 	\begin{align}
 	f_{+}(w)&=(1-w)^{\mu} \\
 	f_{-}(w)&=\alpha w^{\mu}, 
 	\end{align}
 \end{subequations}
	where $\alpha>0$ is the relative strength of depression and $\mu\in[0,1]$ controls the non-linearity of the learning rule.  
	
	The temporal kernels of the STDP rule are normalized: i.e.,  $\int K_{\pm}(\Delta t) d\Delta t = 1$.
 Here, for simplicity,  we assume  that all pairs of pre and post spike times contribute additively to the learning process via \cref{eq:deltaw}.

Empirical studies portray a wide variety of temporal kernels \cite{markram1997regulation,bi1998synaptic,sjostrom2001rate,zhang1998critical,Abbott2000,froemke2006contribution,Nishiyama2000CalciumSR,shouval2002unified,WOODIN2003807}. Specifically, in our work, we used  two families of STDP rules:
	\begin{enumerate*}
		\item  A temporally asymmetric kernel  \cite{bi1998synaptic,zhang1998critical,Abbott2000,froemke2006contribution}.
		\item A temporally symmetric kernel \cite{Nishiyama2000CalciumSR,Abbott2000,shouval2002unified,WOODIN2003807}.
	\end{enumerate*}
	Both of these rules have been observed in the the barrel system of mice, at least for some developmental period \cite{itami2012developmental,itami2016developmental,kimura2019hypothetical}. 
	For the temporally asymmetric kernel we use the exponential model,
	\begin{equation}\label{eq:kernel}
	K_{\pm}(\Delta t)=\frac{e^{\mp  \Delta t/\tau_{\pm}}}{\tau_{\pm}}\Theta(\pm  \Delta t),
	\end{equation} 
	where  $\Theta(x)$ is the Heaviside function, and  $\tau_{\pm}$ are the characteristic timescales of the potentiation $(+)$ or depression $(-)$. We take $\tau_- > \tau_+$ as typically reported. 
	
	For the temporally symmetric learning rule we use a difference of Gaussians model,
	\begin{equation}\label{eq:kernelSymmetric}
	K_{\pm}(\Delta t)=\frac{1}{\tau_\pm \sqrt{2 \pi}} e^{-\frac{1}{2} (\frac{\Delta t}{\tau_\pm})^2},
	\end{equation} 
	where $\tau_\pm$ are the temporal widths. In this case, the order of firing is not important, only the absolute time difference.

	\subsection*{STDP dynamics in the limit of slow learning}  \label{STDP dynamics in the limit of slow learning}
	
 In the limit of a slow learning rate, $\lambda \rightarrow 0$,  we obtain deterministic dynamics for the mean synaptic weights (see \cite{luz2014effect} for a detailed derivation)
	\begin{equation}\label{eq:wdot}
	\frac{\dot{w}_{j}(t)}{\lambda}=I^+_{j}(t) -I^-_{j}(t) 
	\end{equation}
	with
	\begin{equation}\label{eq:wdotpm}
	I^\pm_{j}(t)=f_{\pm}(w_{j}(t)) \int_{-\infty}^{\infty}\Gamma_{j, \text{post}}( \Delta)K_{\pm}(\Delta)d\Delta,
	\end{equation}
	where $\Gamma_{j, \ \text{L4I}}(\Delta)$ is the cross-correlation function between the $j$th thalamic pre-synaptic neuron and the L4I post-synaptic neuron, see detailed derivation in  \nameref{subsec:correl}.

	\subsection*{STDP dynamics of thalamocortical connectivity}
We  simulated the STDP dynamics of 150 upstream thalamic neurons synapsing onto a single L4I neuron in the barrel cortex, see  \nameref{Details of the numeric simulations}.

  \hyperlink{Fig3}{Fig.\ 3a} shows the temporal evolution of the synaptic weights (color coded by their preferred phases).
	As can be seen from the figure, the synaptic weights do not relax to a fixed point; instead there is a continuous remodelling of the entire synaptic population. Examining the order parameters, \hyperlink{Fig3}{Fig.\ 3b and 3c}, reveals that the STDP dynamics converges to a limit cycle.   
	
The continuous remodelling of the synaptic weights causes the preferred phase of the downstream neuron, $\psi_\mathrm{ L4I }$ (see \cref{eq:L4IMeanFiring}) to drift in time,   \hyperlink{Fig3}{Fig.\ 3b} .
	As can be seen from the figure, the drift velocity is not constant. Consequently, the downstream (L4I) neuron `spends' more time in certain phases than others. Thus, the STDP dynamics induce a distribution over time for the preferred phases of the downstream neuron. One can estimate the distribution of the preferred phases of L4I neurons by tracking the phase of a single neuron over time. Alternatively, since our model is purely feed-forward, the preferred phases of different L4I neurons are independent; hence, this distribution can also be estimated by sampling the preferred phases of different L4I neurons at the same time.  \hyperlink{Fig3}{Fig.\ 3d and 3e} show the distribution of preferred phases of thalamic and L4I whisking neurons, respectively. Thus, STDP induces a distribution of preferred phases in the L4I population, which is linked to the temporal distribution of single L4I neurons.

\begin{figure}[tb] \hypertarget{Fig3}{}
	\begin{adjustwidth}{-6cm}{0cm}
		\begin{subfigure}{1\columnwidth}
	\includegraphics[width=0.75\linewidth]{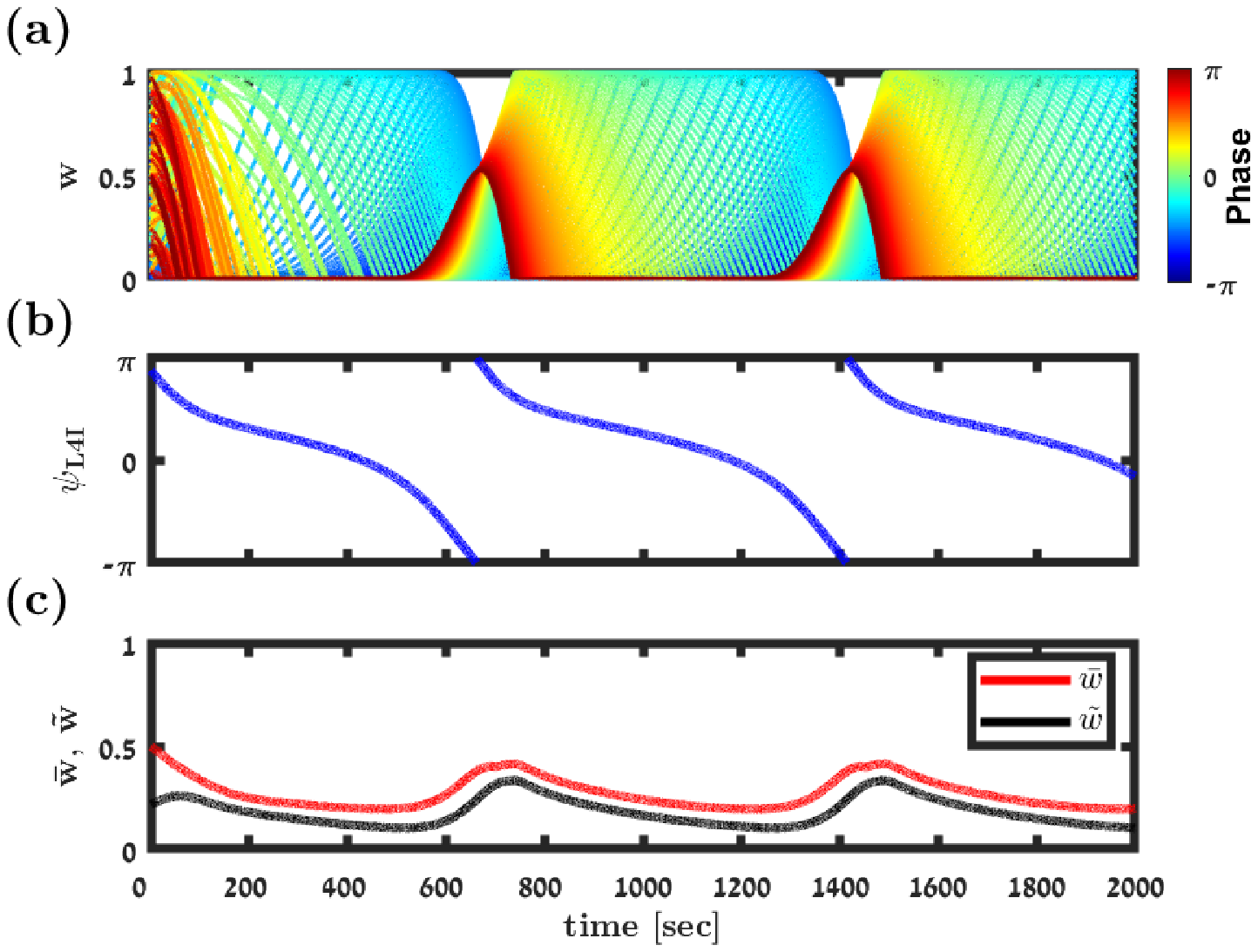}
		\end{subfigure}
			\hspace{-2.5cm}
		\begin{subfigure}{1\columnwidth} 	\vspace{1.5cm}
			\includegraphics[width=0.75\textwidth]{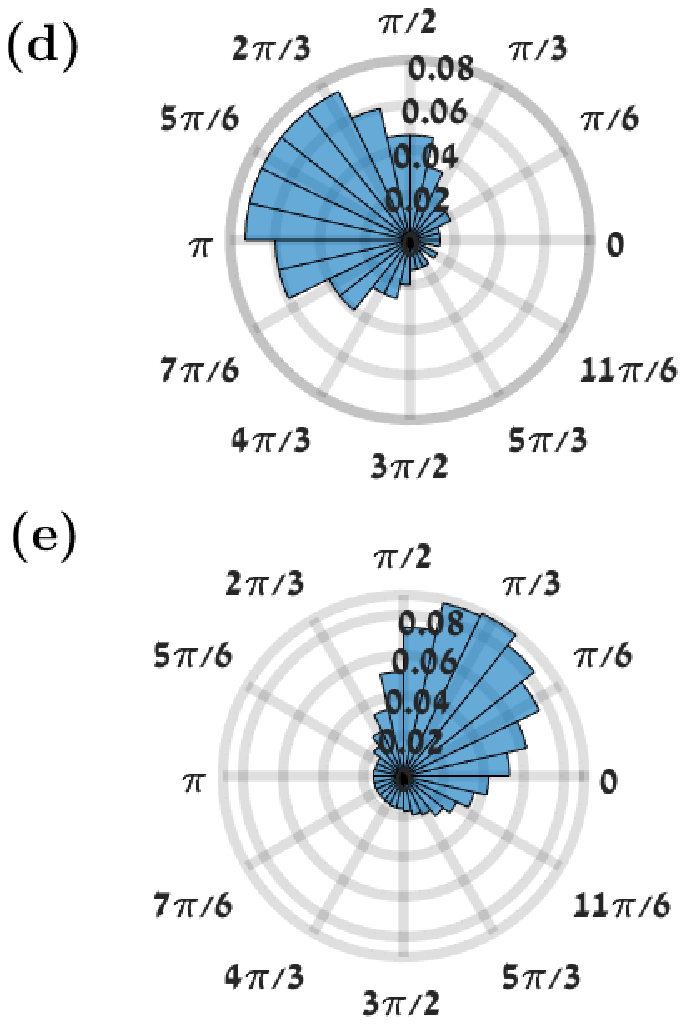}
			\caption{} 
		\end{subfigure} 	
	\caption{ \textbf{Simulation of the STDP dynamics}. (a) Synaptic weights dynamics. Each trace depicts the time evolution of a single synaptic weight, differentiated by color according to its preferred phase, see legend.
	(b) \& (c) Dynamics of the order parameters. The preferred phase of the downstream neuron, $\psi_{ \mathrm{L4I} }$, (in (b)), and the mean, $\bar{w}$, and the magnitude of the first Fourier component, $\tilde{w}$, (in red and black, respectively, in (c)) are shown as a function of time. 
	(d) The distribution of preferred phases in the thalamic population that served as input to the downstream L4I neuron is presented as a polar histogram. The distribution followed \cref{eq:VM} with $\kappa_{ \mathrm{VPM} } = 1$ and $\psi_{ \mathrm{VPM} } = 5 \pi /6 \ \text{rad}$. 
	(e) The temporal distribution of preferred phases of the downstream L4I neuron is shown in a polar plot. Fitting the von-Mises distribution yielded  $\kappa_\mathrm{L4I}=1.1$ and $\psi_{ \mathrm{L4I}}=0.8 \  \text{rad}$.
		In this simulation  we used the following parameters: $N = 150$,  $\bar{\nu}=\nu/(2 \pi)= 7\text{hz}$, $\gamma_{ \mathrm{VPM} }=1$, $D_{ \mathrm{VPM} } = 10 \text{hz}$, and $d=3 \text{ms}$. The temporally asymmetric STDP rule, \cref{eq:kernel}, was used with $\tau_-=50\text{ms}$, $\tau_+=22\text{ms}$, $\mu=0.01$, $\alpha=1.1$, and $\lambda=0.01$.
		}
	\label{fig:Fig3}
		\end{adjustwidth}
\end{figure}

\begin{figure*}[tb!]
		\begin{adjustwidth}{-6.2cm}{0cm}
			
			\centering

					\begin{subfigure}[t]{0.007\textwidth}
					\textbf{(a)} 
				\end{subfigure}
				\begin{subfigure}[t]{0.42\textwidth}  \vspace{0.007\textwidth}
					\includegraphics[width=\linewidth, valign=t]{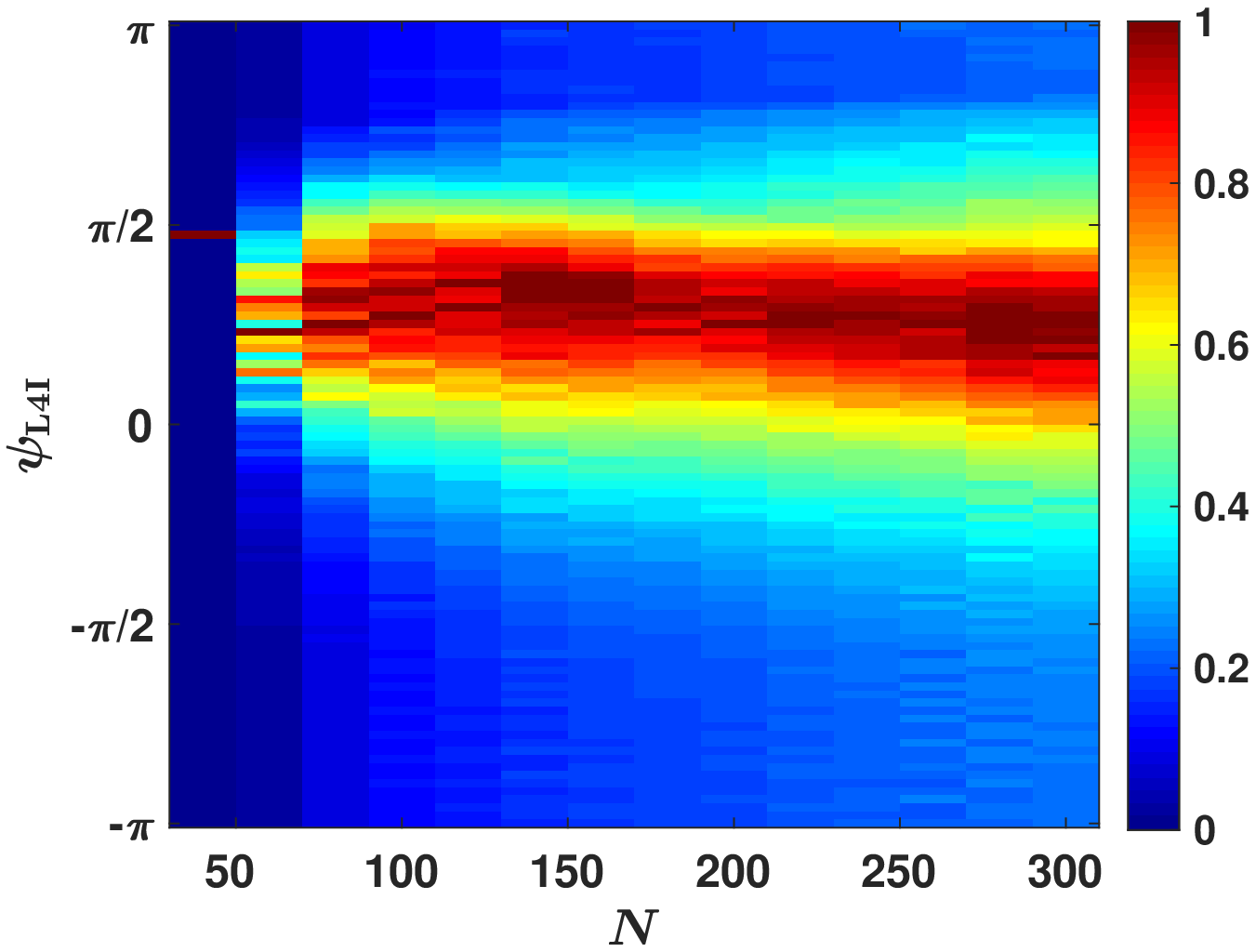}\\[3pt]
					\caption{}
					\label{fig:Fig4a}
				\end{subfigure}\hspace{0.5cm} \vspace{0.02\textwidth}
				\begin{subfigure}[t]{0.007\textwidth}
				\textbf{(b)} 
			\end{subfigure}
			\begin{subfigure}[t]{0.42\textwidth}  \vspace{0.007\textwidth}
				\includegraphics[width=\linewidth, valign=t]{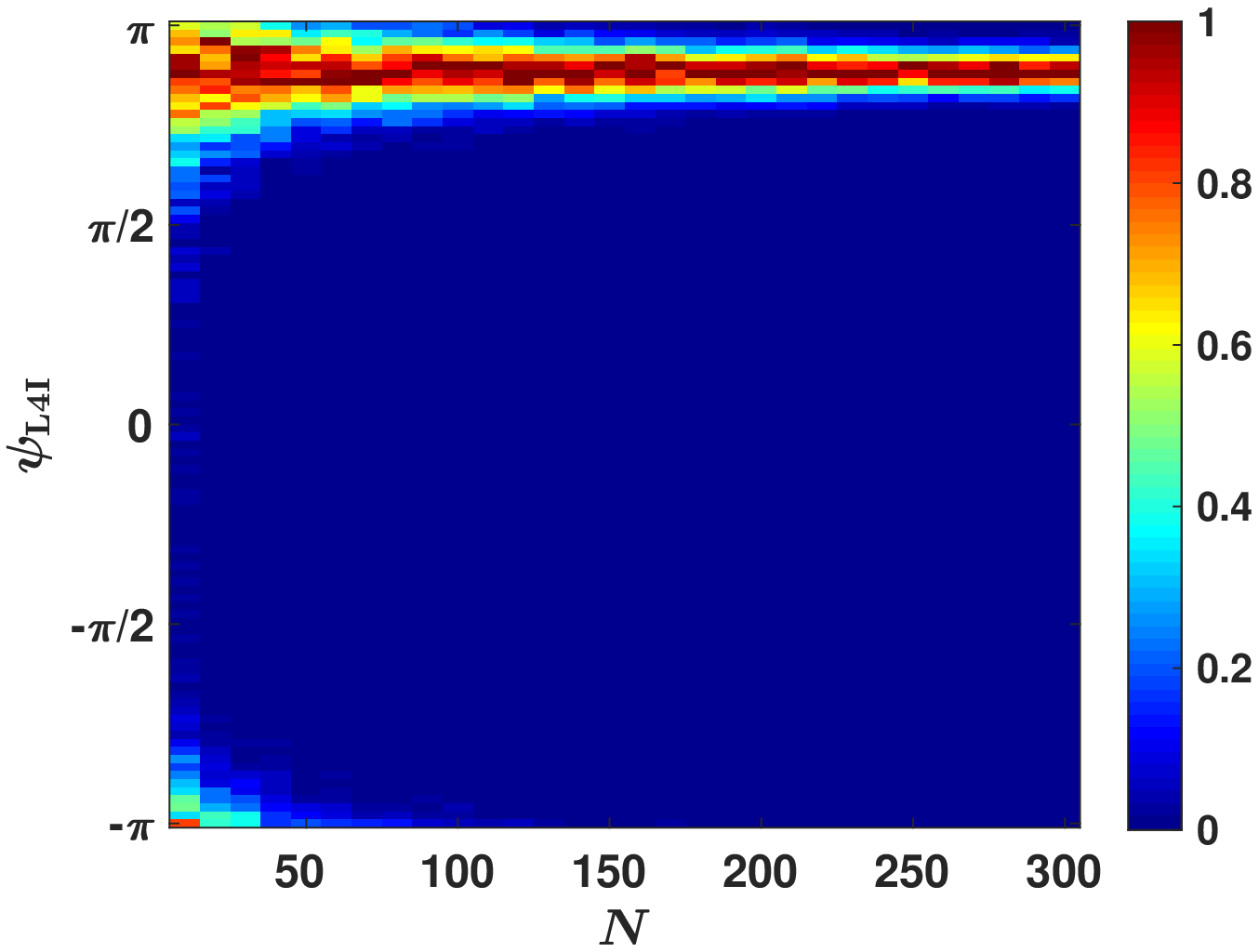}\\[3pt]
				\caption{}
				\label{fig:Fig4b}
			\end{subfigure}\hspace{0.5cm} \vspace{0.02\textwidth}
				\begin{subfigure}[t]{0.01\textwidth}
					\textbf{(c)}
				\end{subfigure}
				\begin{subfigure}[t]{0.42\textwidth}  \vspace{0.007\textwidth} 	
					\includegraphics[width=\linewidth, valign=t]{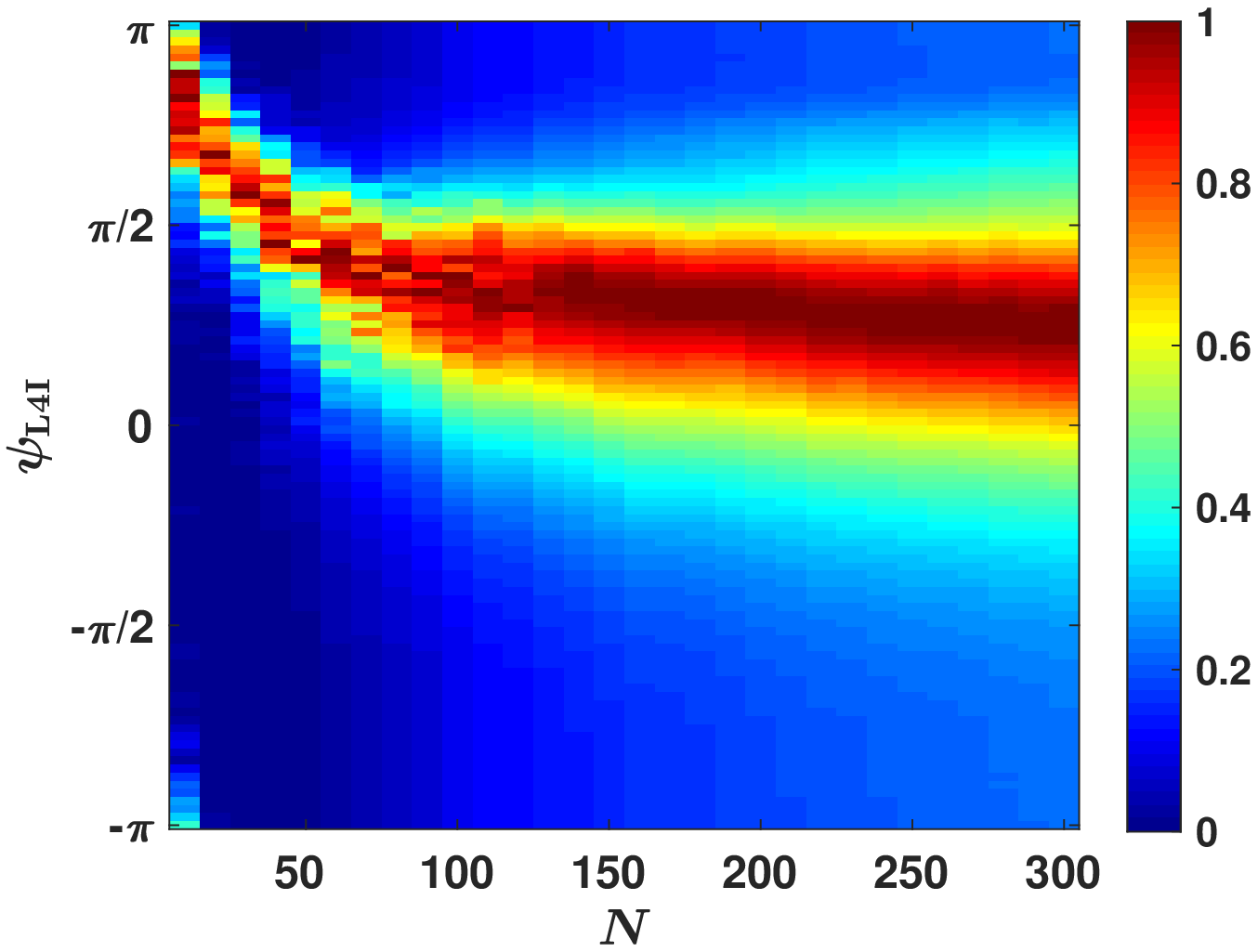}\\[3pt]
					\caption{}
					\label{fig:Fig4c}
				\end{subfigure}\hfill \vspace{0.0001\textwidth}
			
			\caption{ \textbf{The effects of population size.} The distribution of the preferred phases of L4I neurons, $\psi_\mathrm{L4I}$, is shown as a function of $N$ for different sources of variability. In  each column (value of $N$), the non-normalized distribution: $\Pr (\psi) / \max \{ \Pr (\psi) \} $, is presented by color. 
			(a) The temporal distribution of the preferred phase of a single L4I neuron as a result of STDP dynamics without quenched disorder or averaging (see \nameref{Details of the phase distribution}) is presented.  
			(b) Distribution due to quenched disorder - without STDP. The distribution of L4I neurons in the randomly pooling model was estimated from 1000 trials of drawing the preferred phases of $N$ VPM neurons in an $iid$ manner from \cref{eq:VM}. 
			(c)  The distribution due to both quenched disorder and STDP dynamics is shown.  Unless stated otherwise, the parameters used in these graphs are as follows: $\bar{\nu}=\nu/(2 \pi)= 7\text{Hz}$, $\kappa_\mathrm{VPM}=1$, $\gamma=0.9$, $D = 10 \text{hz}$,  $d=3 \text{ms}$, $\phi_0= 5 \pi/6$. For the STDP we used the temporally exponential asymmetric kernel, \cref{eq:kernel}, with $\tau_-=50\text{ms}$, $\tau_+=22\text{ms}$, $\mu=0.01$, $\alpha=1.1$, and $\lambda=0.01$.}  
			\label{fig:Fig4}
		\end{adjustwidth}
	\end{figure*}

\cref{fig:Fig4a} shows the (non-normalized) distribution of preferred phases induced by STDP as a function of the number of pooled thalamic neurons, $N$. For large $N$, STDP dynamics converge to the continuum limit of \cref{eq:wdot}, and the distribution converges to a limit that is independent of $N$.
This is in contrast with the simple feed-forward pooling model lacking plasticity (cf.\ \cref{fig:Fig4b}). In this model, the distribution of preferred phases in the cortical population results from a random process of pooling phases from the thalamic population. We shall refer to this type of variability as quenced disorder, since this randomness is frozen and does not fluctuate over time. This process is characterized by a distribution of preferred phases with vanishing width, $\kappa_{ \mathrm{L4I}} \rightarrow \infty$, in the large $N$ limit, \cref{fig:Fig4b}. In addition to the different widths of the distribution, the mean preferred phases also differs greatly. Whereas in \cref{fig:Fig4a} the mean preferred phase is determined by the STDP rule, without STDP, \cref{fig:Fig4b}, the mean phases of L4I neurons is given by the mean preferred phase of the VPM population shifted by the delay, $\psi_{ \mathrm{L4I} } = \psi_{ \mathrm{VPM} } + d \nu$.

For small $N$, $N \lessapprox 70$, STDP induces a point measure distribution over time, \cref{fig:Fig4a}. This is due to pinning in the noiseless STDP dynamics in the mean field limit, $\lambda \rightarrow 0$. Stochastic dynamics, due to noisy neuronal responses, could overcome pinning. \cref{fig:Fig4c} depicts the distribution of the preferred phases for different values of $N$, which results from both STDP dynamics and quenched averaging over different realizations for the preferred phases of the thalamic neurons. For small values of $N$, the distribution is dominated by the quenched statistics, in terms of its narrow width and mean that is dominated by the mean phase of the thalmic population. As $N$ increases, the distribution widens and is centered around a preferred phase that is determined by the STDP. Thus, activity dependent plasticity helps to shape the distribution of preferred phases in the downstream population. Below, we study how these different parameters affect the ways in which  STDP shapes the distribution; hence, we will not average over the quenched disorder.

\begin{figure*}[tb!]
		\begin{adjustwidth}{-6.2cm}{0cm}
			
			\centering

					\begin{subfigure}[t]{0.007\textwidth}
					\textbf{(a)} 
				\end{subfigure}
				\begin{subfigure}[t]{0.42\textwidth}  \vspace{0.007\textwidth}
					\includegraphics[width=\linewidth, valign=t]{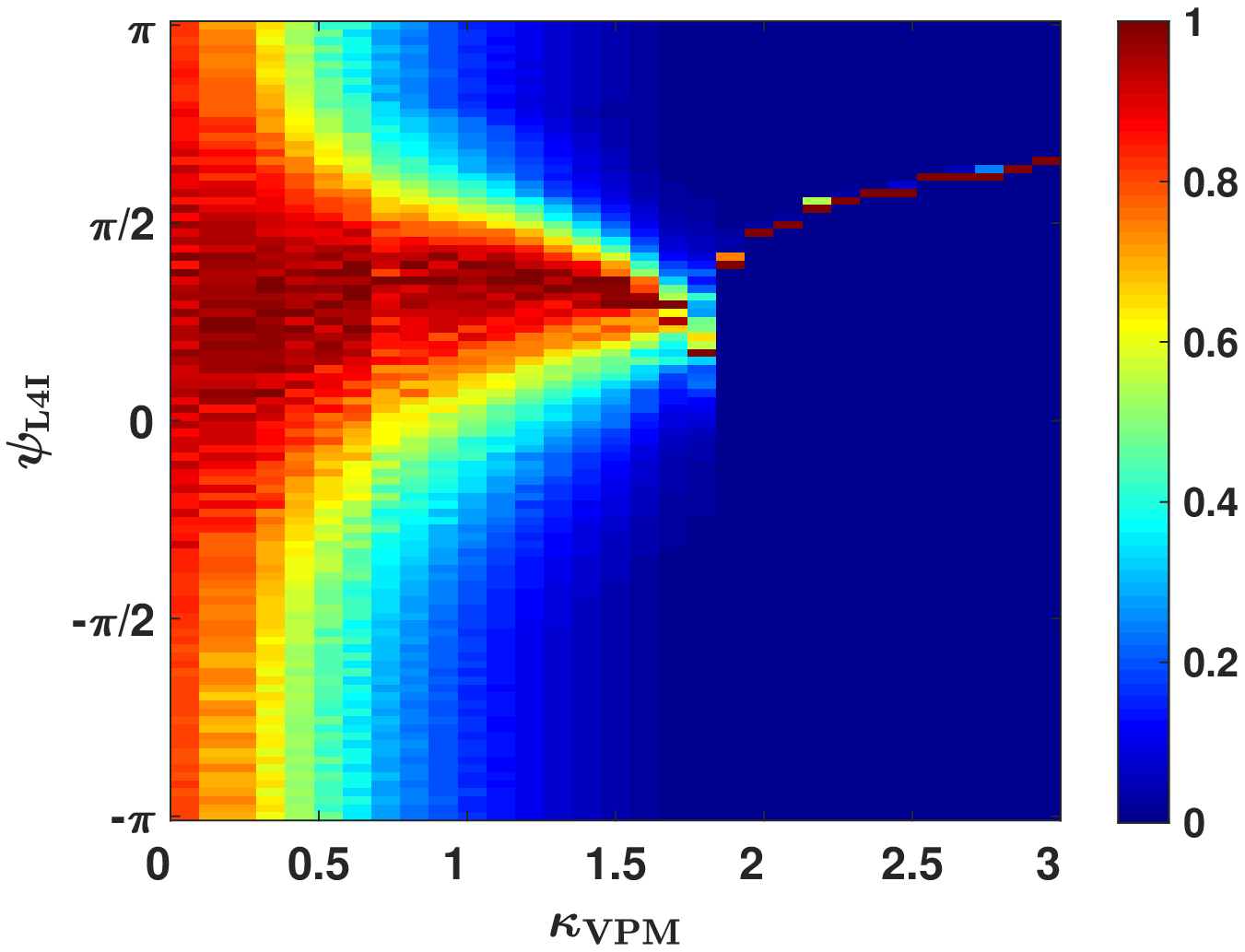}\\[3pt]
					\caption{}
					\label{fig:Fig5a}
				\end{subfigure}\hspace{0.5cm} \vspace{0.02\textwidth}
				\begin{subfigure}[t]{0.007\textwidth}
				\textbf{(b)} 
			\end{subfigure}
			\begin{subfigure}[t]{0.42\textwidth}  \vspace{0.007\textwidth}
				\includegraphics[width=\linewidth, valign=t]{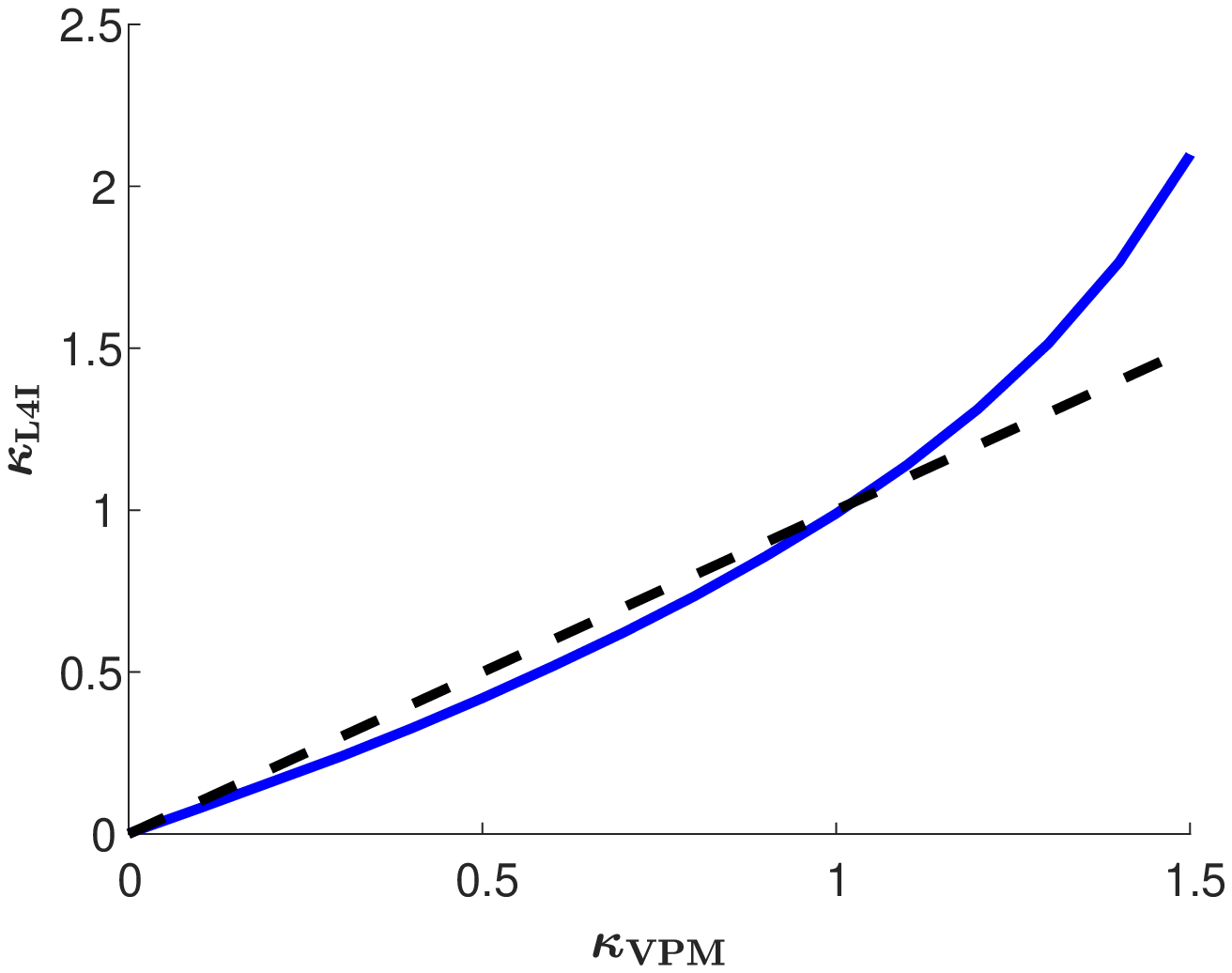}\\[3pt]
				\caption{}
				\label{fig:Fig5b}
			\end{subfigure}\hspace{0.5cm} \vspace{0.02\textwidth}
				\begin{subfigure}[t]{0.01\textwidth}
					\textbf{(c)}
				\end{subfigure}
				\begin{subfigure}[t]{0.42\textwidth}  \vspace{0.007\textwidth} 	
					\includegraphics[width=\linewidth, valign=t]{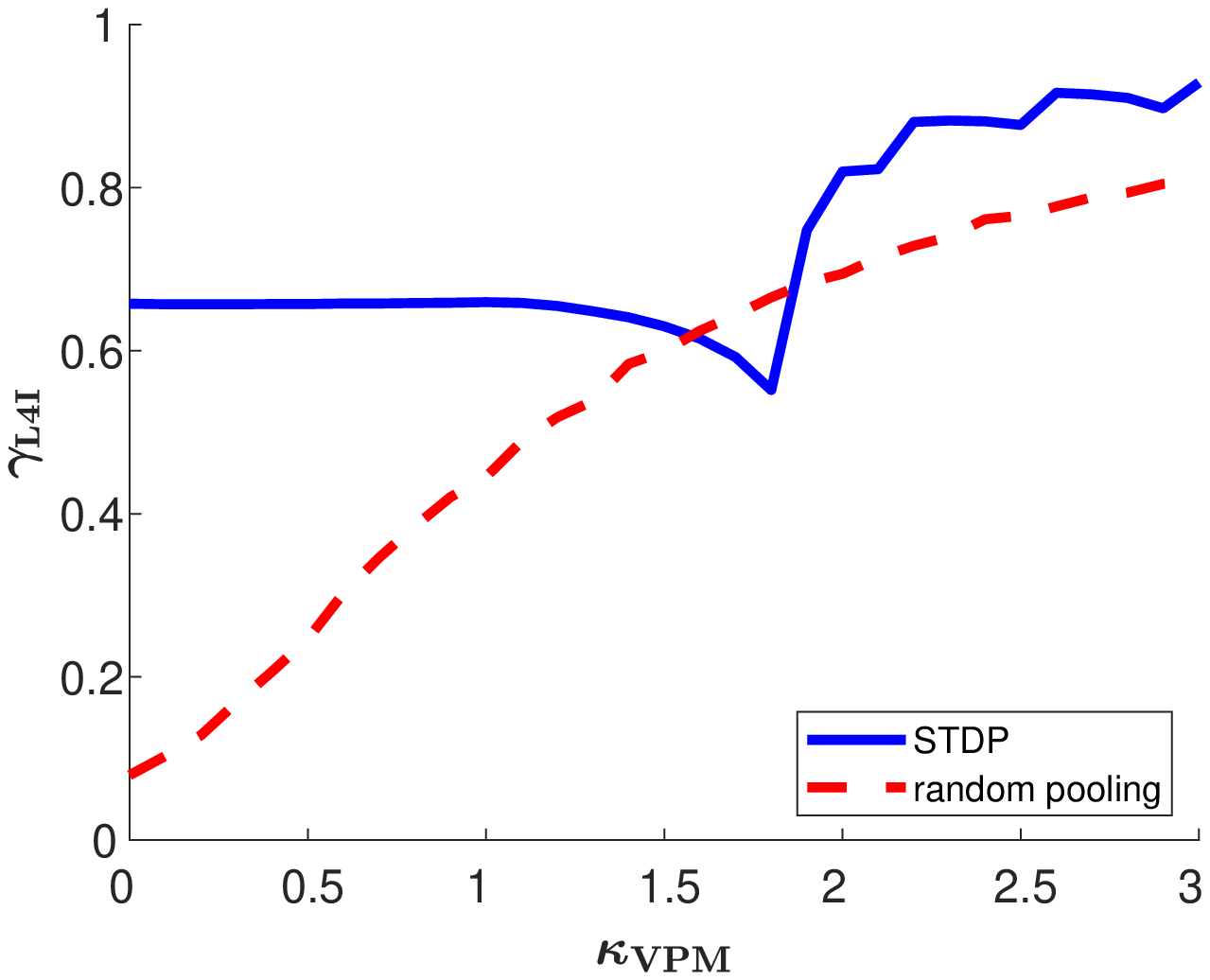}\\[3pt]
					\caption{}
					\label{fig:Fig5c}
				\end{subfigure}\hfill \vspace{0.0001\textwidth}
			
			\caption{ \textbf{The effect of the distribution width of the upstream population, $\kappa_{ \mathrm{VPM}}$.} 
			(a) The (non-normalized) distribution of the preferred phases of L4I neurons, $\psi_{ \mathrm{L4I}}$, is shown by color as a function of the width of the distribution of preferred phases in VPM, $\kappa_{ \mathrm{VPM}}$.  
			(b) The distribution width in layer 4, $\kappa_{ \mathrm{L4I}}$, is shown as a function of $\kappa_{ \mathrm{VPM}}$ (blue). The identity line is shown  (dashed black) for comparison. 
			(c) The modulation depth in the downstream population, $\gamma_{ \mathrm{L4I}}$, is shown as a function of  $\kappa_{ \mathrm{VPM}}$ (blue). For comparison  the modulation depth of the uniform pooling model is also presented (dashed red). The parameters used here are: $\bar{\nu}=\nu/(2 \pi)= 7\text{Hz}$, $\kappa_\mathrm{VPM}=1$, $\gamma=1$, $D = 10 \text{hz}$, $d=3 \text{ms}$, $\phi_0= 5 \pi/6$. For the STDP we used the temporally exponential asymmetric kernel, \cref{eq:kernel}, with
			$\tau_-=50\text{ms}$, $\tau_+=22\text{ms}$, $\mu=0.01$, $\alpha=1.1$, and $\lambda=0.01$.}  
			\label{fig:Fig5}
		\end{adjustwidth}
	\end{figure*}

\subparagraph{Parameters characterizing the upstream input.}
\cref{fig:Fig5a} depicts the distribution of preferred phases as a function of the distribution width of their thalamic input, $\kappa_{ \mathrm{VPM} }$. For a uniform input distribution, $\kappa_{ \mathrm{VPM}} = 0$, the downstream distribution is also uniform, $\kappa_{ \mathrm{L4I}} = 0$, see \cite{luz2016oscillations}. As the distribution in the VPM becomes narrower, so does the distribution in the L4I population. If the distribution of the thalamic population is narrower than a certain critical value, STDP will converge to a fixed point and $ \kappa_{ \mathrm{L4I}} $ will diverge. Typically, we find that the width of the cortical distribution, $ \kappa_{ \mathrm{L4I}} $, is similar to or larger than that of the upstream distribution, $  \kappa_{ \mathrm{VPM} }$,  \cref{fig:Fig5b}. This sharpening is obtained via STDP by selectively amplifying certain phases while attenuating others. Consequently the rhythmic component, in terms of the modulation depth, $\gamma_{ \mathrm{L4I}}$, is also typically amplified relative to the uniform pooling model, \cref{fig:Fig5c}.

\begin{figure*}[tb!]
			
			\centering

					\begin{subfigure}[t]{0.007\textwidth}
					\textbf{(a)} 
				\end{subfigure}
				\begin{subfigure}[t]{0.42\textwidth}  \vspace{0.007\textwidth}
					\includegraphics[width=\linewidth, valign=t]{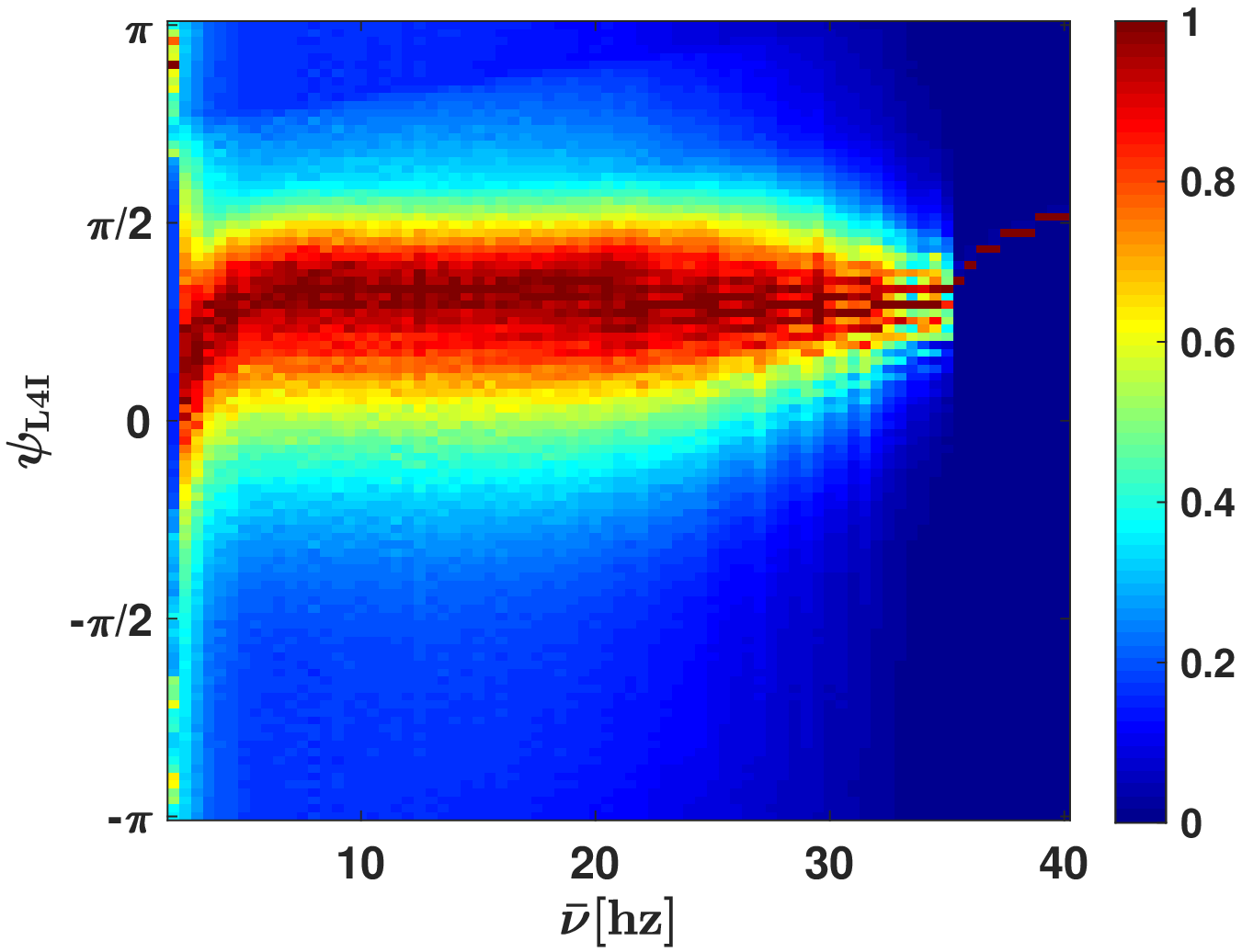}\\[3pt]
					\caption{}
					\label{fig:Fig6a}
				\end{subfigure}\hspace{0.5cm} \vspace{0.02\textwidth}
				\begin{subfigure}[t]{0.007\textwidth}
				\textbf{(b)} 
			\end{subfigure}
			\begin{subfigure}[t]{0.42\textwidth}  \vspace{0.007\textwidth}
				\includegraphics[width=\linewidth, valign=t]{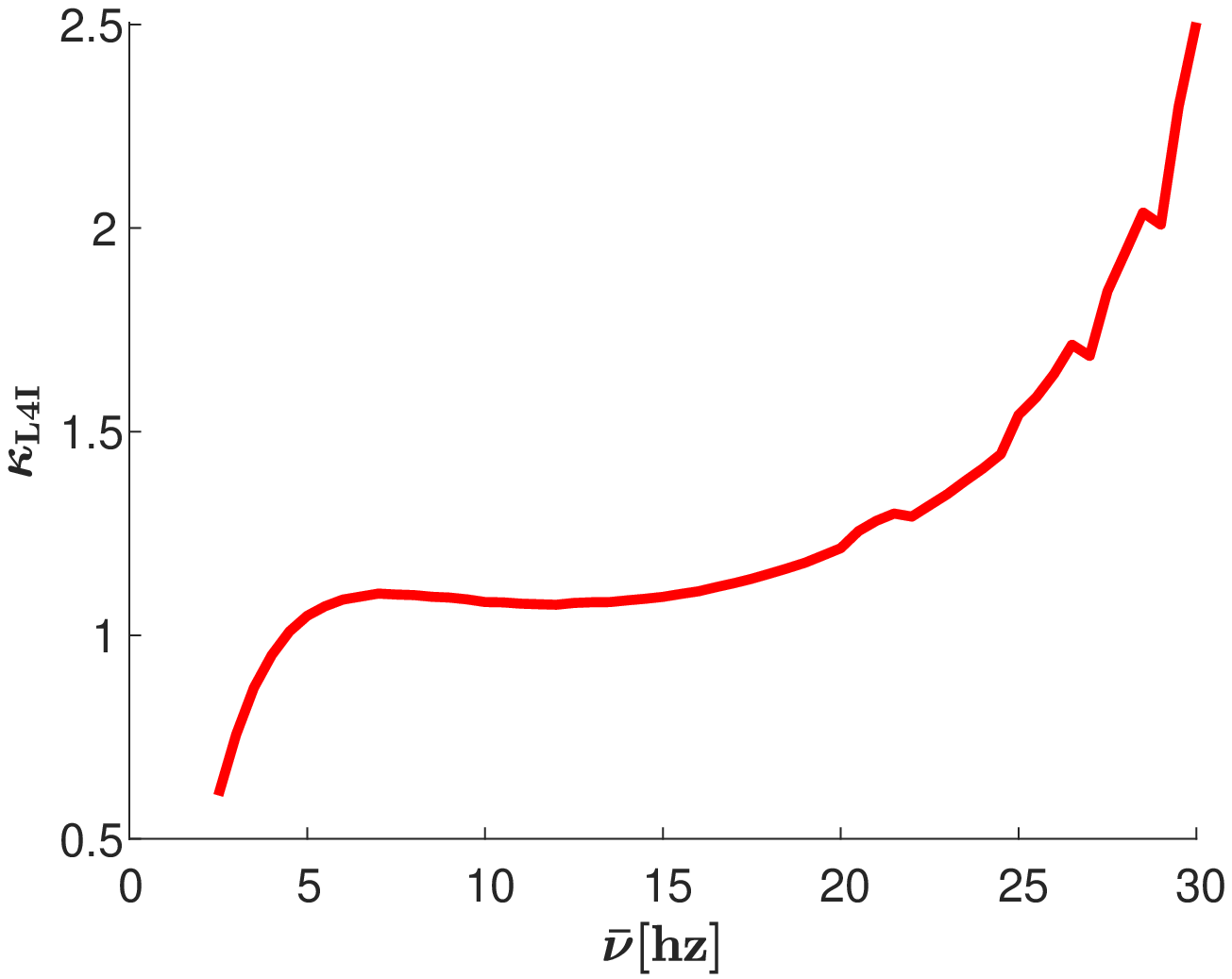}\\[3pt]
				\caption{}
				\label{fig:Fig6b}
			\end{subfigure}\hspace{0.5cm} \vspace{0.02\textwidth}
			\hfill \vspace{0.0001\textwidth}
			
			\caption{ \textbf{Effects of whisking frequency.} 
			(a) The (non-normalized) distribution of L4I neuron phases, $\psi_{ \mathrm{L4I}}$, is depicted by color as a function, $\bar{\nu}=\nu/(2 \pi)$. 
			(b)  The width of the distribution, $\kappa_{\mathrm{L4I}}$,  of L4I neurons, is shown as a function of $\bar{\nu}$. 
			The parameters used in these graphs were: $\kappa_{ \mathrm{VPM}}=1$, $\psi_{ \mathrm{VPM}}= 5 \pi/6$ $\gamma=0.9$, $D = 10 \text{hz}$, and $d=3 \text{ms}$.
			We used the temporally asymmetric exponential learning rule, \cref{eq:kernel}, with $\tau_-=50\text{ms}$, $\tau_+=22\text{ms}$, $\mu=0.01$, $\alpha=1.1$, and $\lambda=0.01$.}  
			\label{fig:Fig6}
	\end{figure*}

\begin{figure*}[tb!]
			
			\centering

					\begin{subfigure}[t]{0.007\textwidth}
					\textbf{(a)} 
				\end{subfigure}
				\begin{subfigure}[t]{0.42\textwidth}  \vspace{0.007\textwidth}
					\includegraphics[width=\linewidth, valign=t]{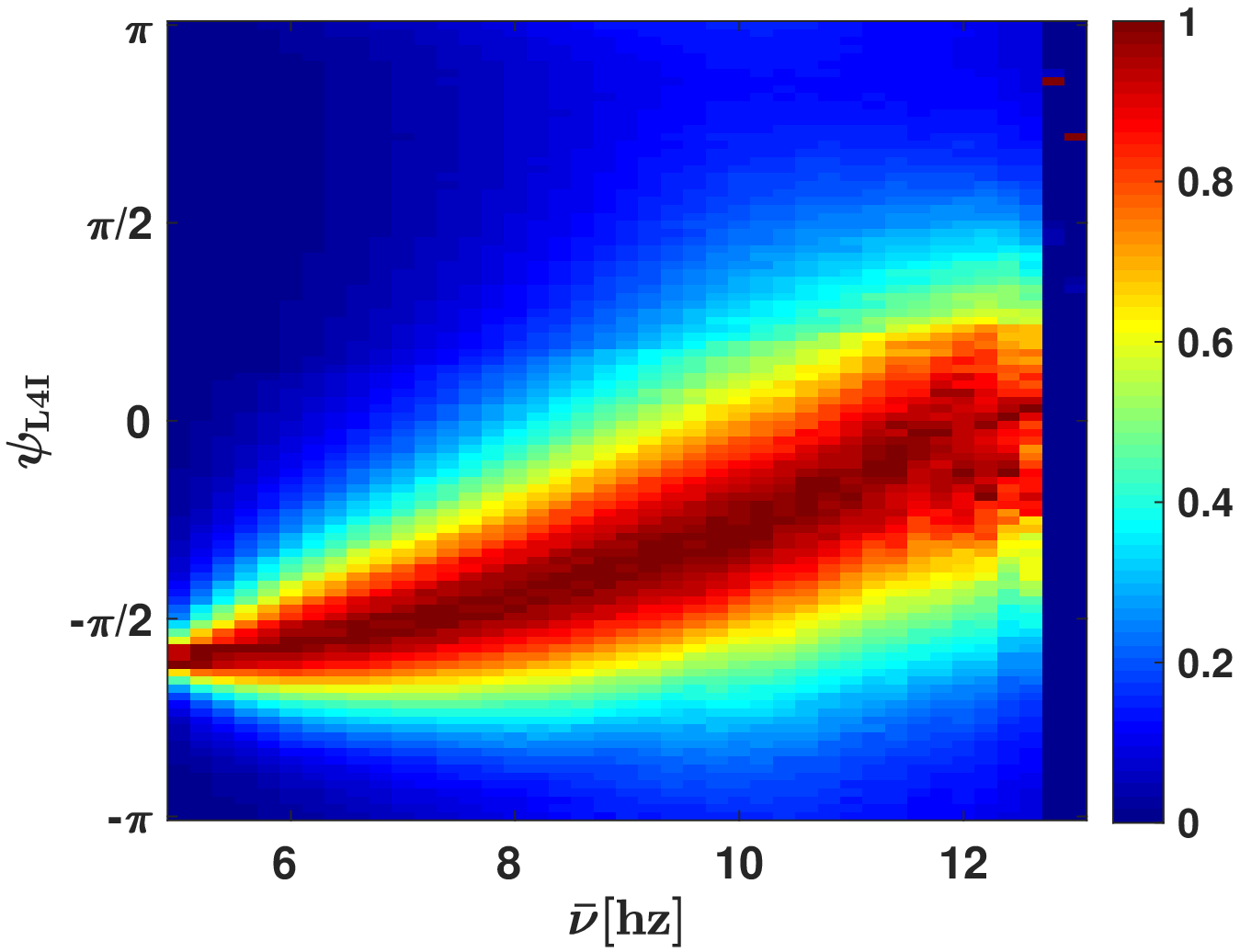}\\[3pt]
					\caption{}
					\label{fig:Fig7a}
				\end{subfigure}\hspace{0.5cm} \vspace{0.02\textwidth}
				\begin{subfigure}[t]{0.007\textwidth}
				\textbf{(b)} 
			\end{subfigure}
			\begin{subfigure}[t]{0.42\textwidth}  \vspace{0.007\textwidth}
				\includegraphics[width=\linewidth, valign=t]{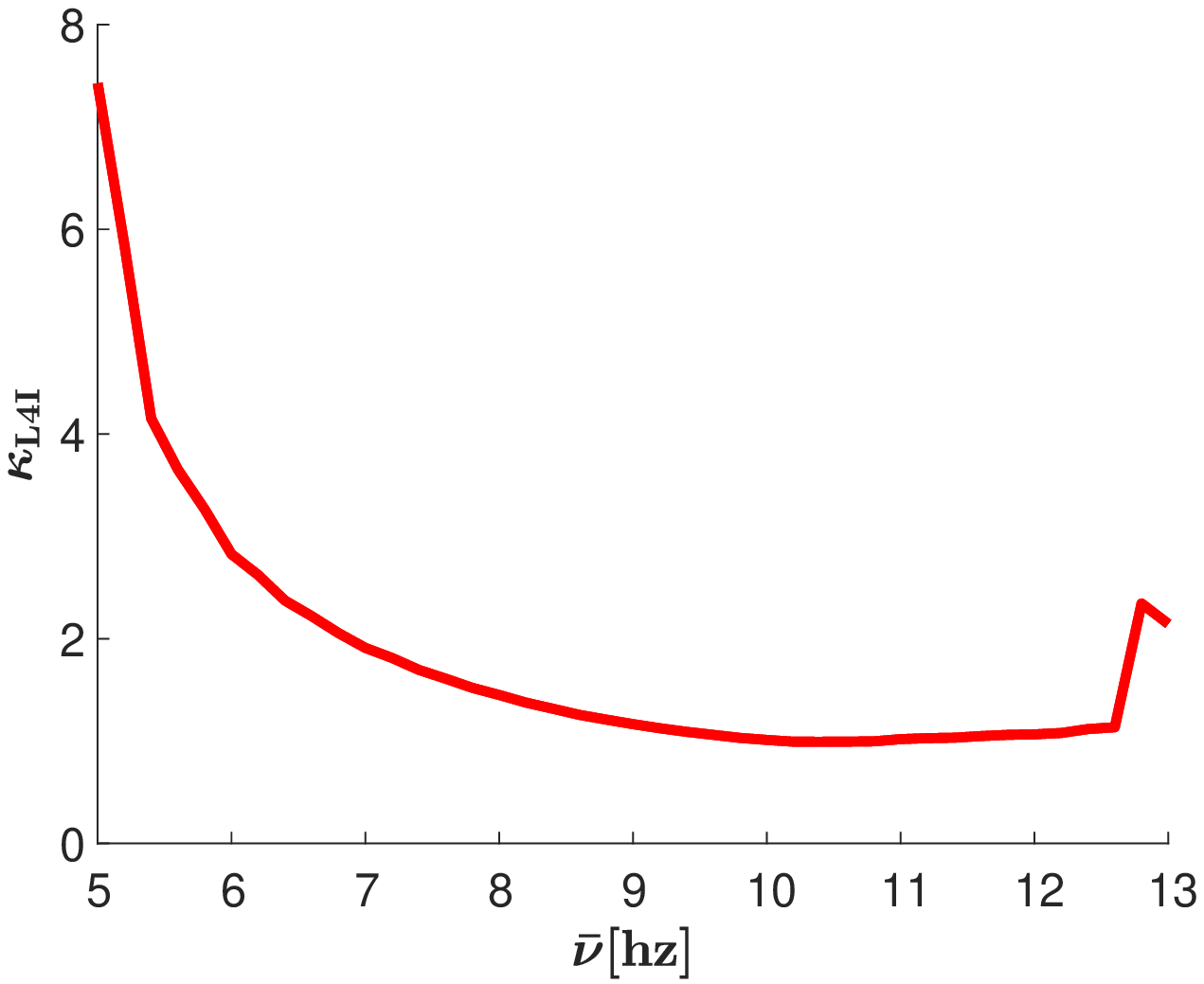}\\[3pt]
				\caption{}
				\label{fig:Fig7b}
			\end{subfigure}\hspace{0.5cm} \vspace{0.02\textwidth}
			\hfill \vspace{0.0001\textwidth}
			
			\caption{ \textbf{Effects of whisking frequency - the temporally symmetric STDP rule.} 
			(a) The (non-normalized) distribution of L4I neuron phases, $\psi_{ \mathrm{L4I}}$, is depicted by color as a function, $\bar{\nu}=\nu/(2 \pi)$. 
			(b) The width of the distribution, $\kappa_{\mathrm{L4I}}$, of L4I neurons, is shown as a function of $\bar{\nu}$.
			Unless stated otherwise, the parameters used here were: $\kappa_\mathrm{VPM}=1$, $\psi_{ \mathrm{VPM}}= 5 \pi/6$, $\gamma=0.9$, $D = 10 \text{hz}$, and $d=10 \text{ms}$. We  used the temporally asymmetric exponential learning rule, \cref{eq:kernelSymmetric}, with: $\tau_-=50\text{ms}$, $\tau_+=22\text{ms}$, $\mu=0.01$, $\alpha=1.1$, and $\lambda=0.01$.}  
			\label{fig:Fig7}
	\end{figure*}

The effect of the whisking frequency is shown in \cref{fig:Fig6a}. For moderate rhythms that are on a similar timescale to that of the STDP rule, STDP dynamics can generate a wide distribution of preferred phases. However, in the high frequency limit, $\nu \rightarrow \infty$, the synaptic weights converge to a uniform solution with $w(\phi) =(1+\alpha^{1/\mu})^{-1}$, $\forall \phi $ (see \cite{luz2016oscillations}). In this limit, due to the uniform pooling, there is no selective amplification of phases and the whisking signal is transmitted downstream due to the selectivity of the thalamic population, $ \kappa_{ \mathrm{VPM} } >0$. Consequently, at high frequencies, the distribution of preferred phases will be extremely narrow, $ \mathcal{O} (1/\sqrt{N})$ - due to the quenched disorder, and the rhythmic signal will be attenuated, $\gamma_{ \mathrm{L4I}} = \gamma_{ \mathrm{VPM}} \times I_1 (\kappa) /I_0 ( \kappa) $ (where $I_j ( \kappa)$ is the modified Bessel function of order $j$), \cref{fig:Fig6b}. The rate of convergence to the high frequency limit is governed by the smoothness of the STDP rule. Discontinuity in the STPD rule, such as in our choice of a temporally asymmetric rule, will induce algebraic convergence in $\nu$ to the high frequency limit, whereas a smooth rule, such as our choice of a temporally symmetric rule, will manifest in exponential convergence, compare \cref{fig:Fig6} with \cref{fig:Fig7},  see also \cite{luz2016oscillations}. 
In our choice of parameters, $\tau_+ \approx 20 $ms and $\tau_- \approx 50$ms, STDP dynamics induced a wide distribution for frequencies in the $\alpha, \beta$ and low $\gamma$ bands in the case of asymmetric learning rule \cref{fig:Fig6}, and around $  5$ - $15  \text{hz}$ in the case of the symmetric learning rule  \cref{fig:Fig7}.

\begin{figure} \hypertarget{Fig8}{}
	\begin{adjustwidth}{-6cm}{0cm}
		\begin{subfigure}{0.65\columnwidth}
			\includegraphics[width=\textwidth]{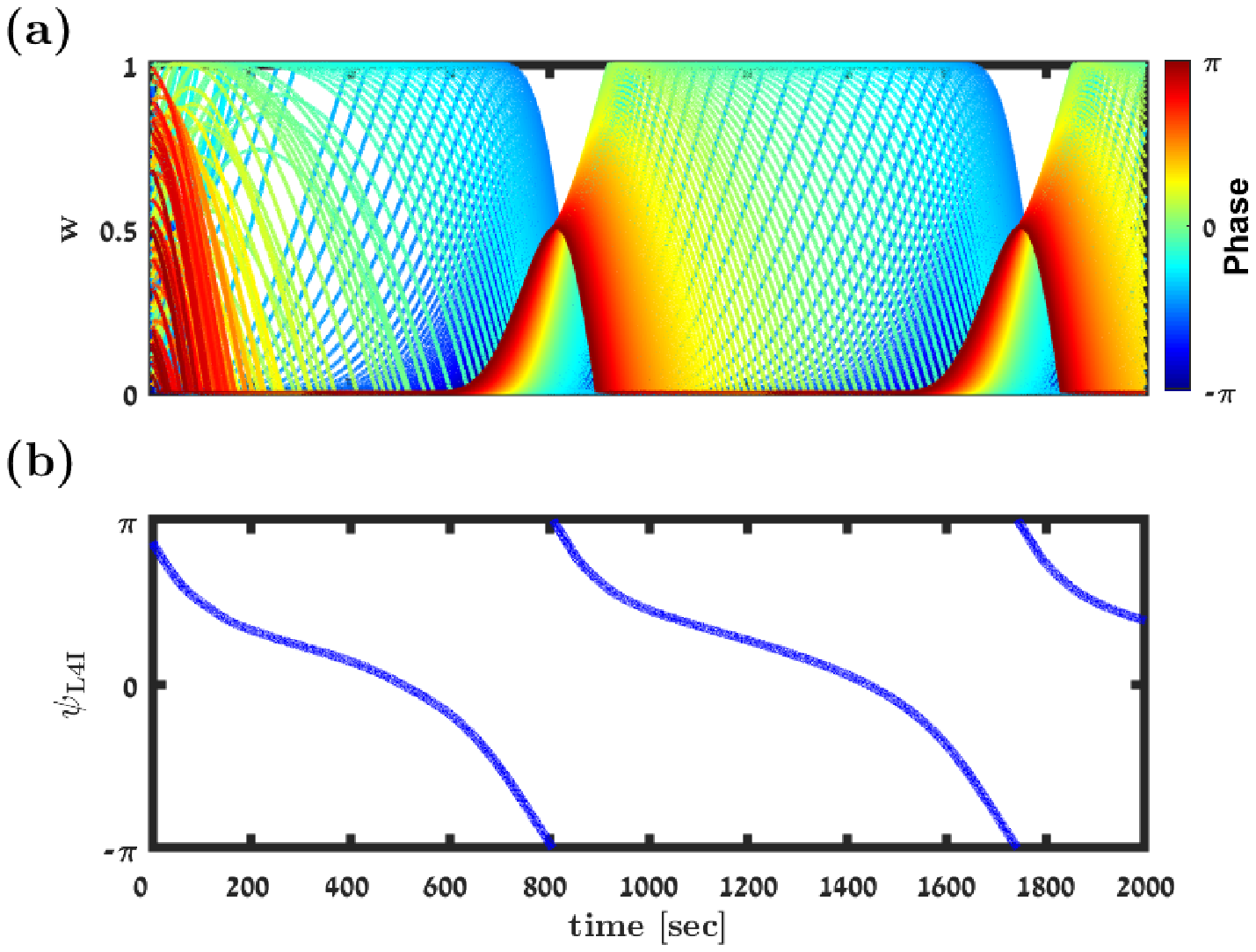}
			\caption{}
		\end{subfigure}
		\hfill
		\begin{subfigure}{0.65\columnwidth}
			\includegraphics[width=\textwidth]{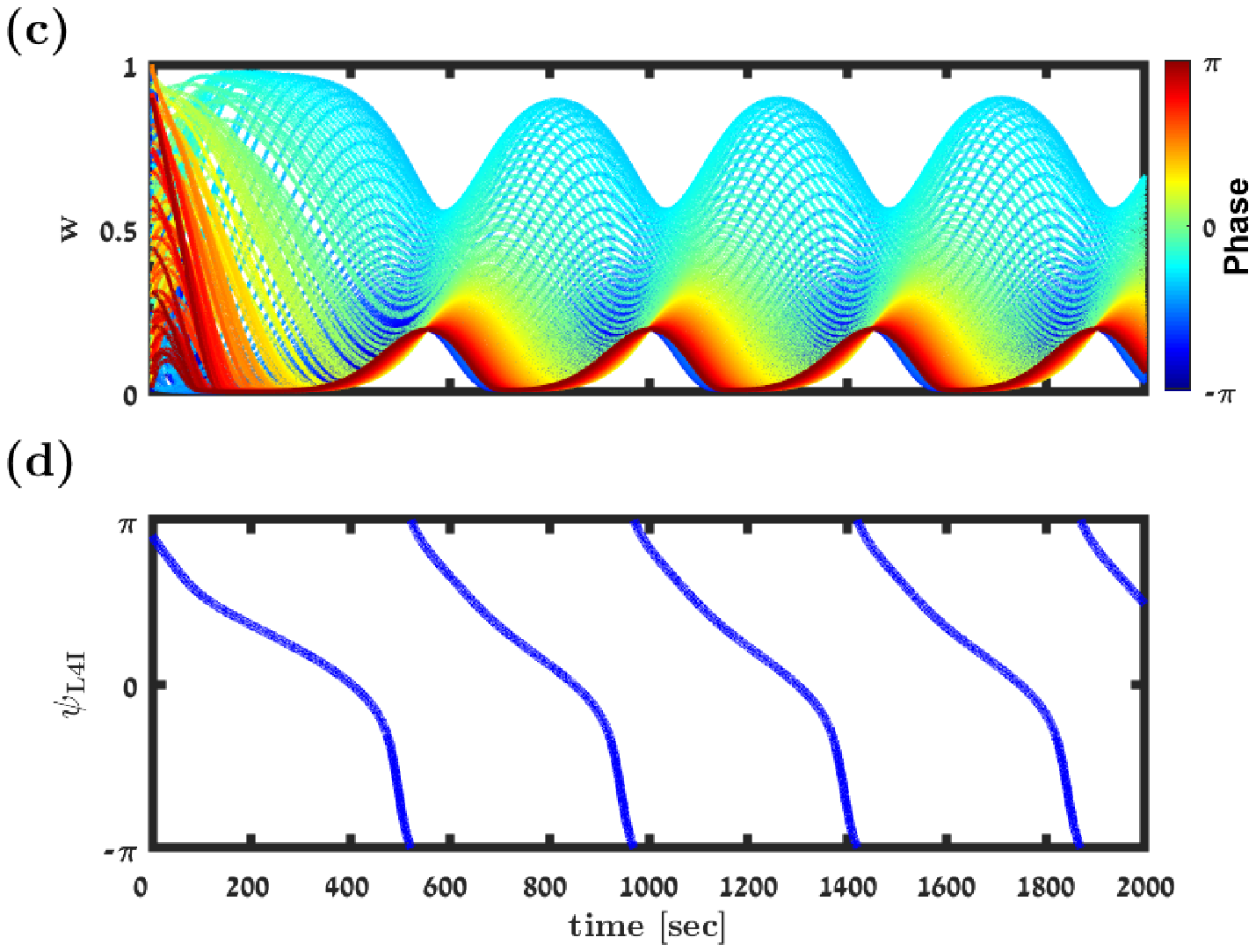}
			\caption{} 
		\end{subfigure} 
		\begin{subfigure}{0.65\columnwidth} 
			\includegraphics[width=\textwidth]{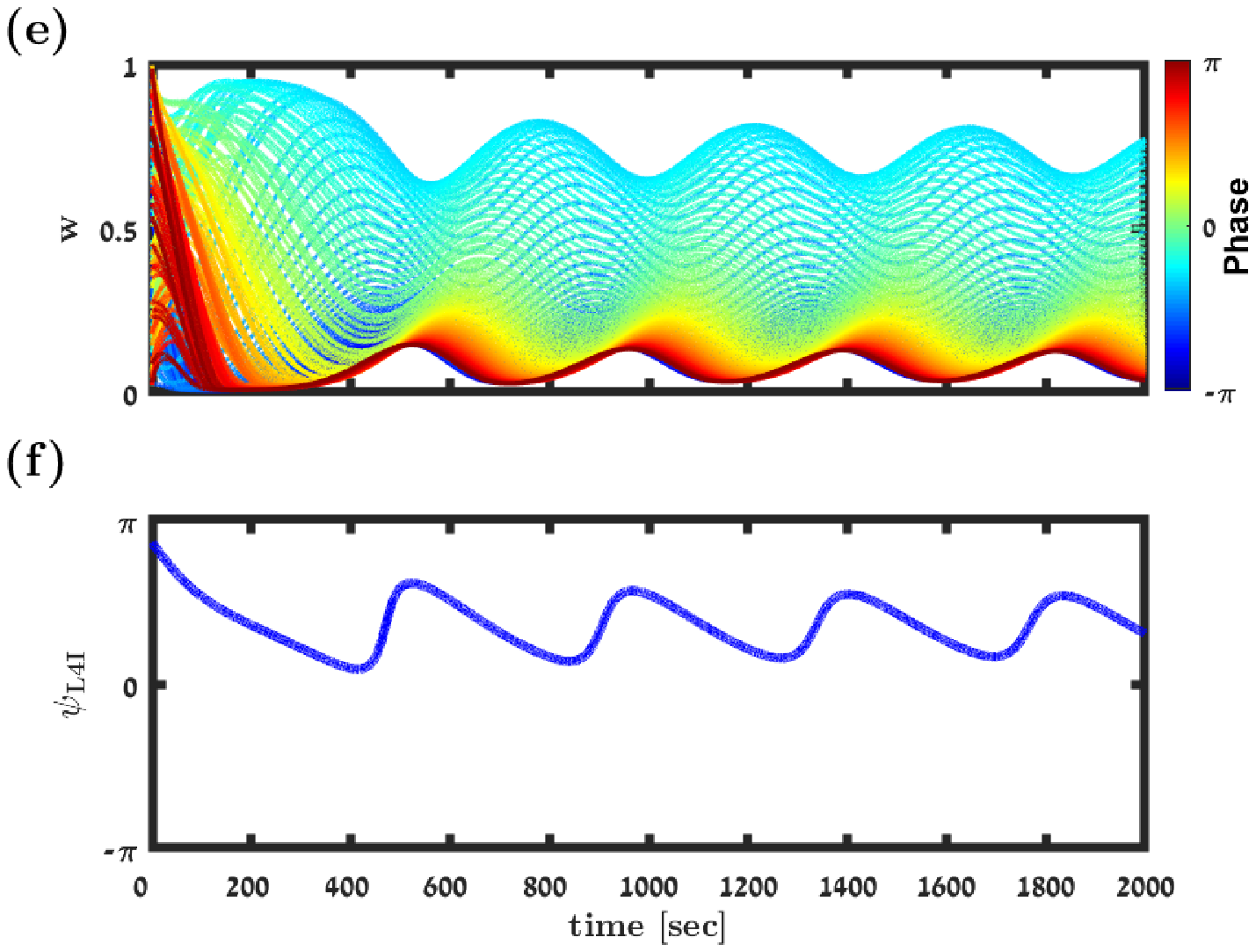} 
			\caption{} 
		\end{subfigure}  
		\hfill 
		\begin{subfigure}{0.65\columnwidth} 
			\includegraphics[width=\textwidth]{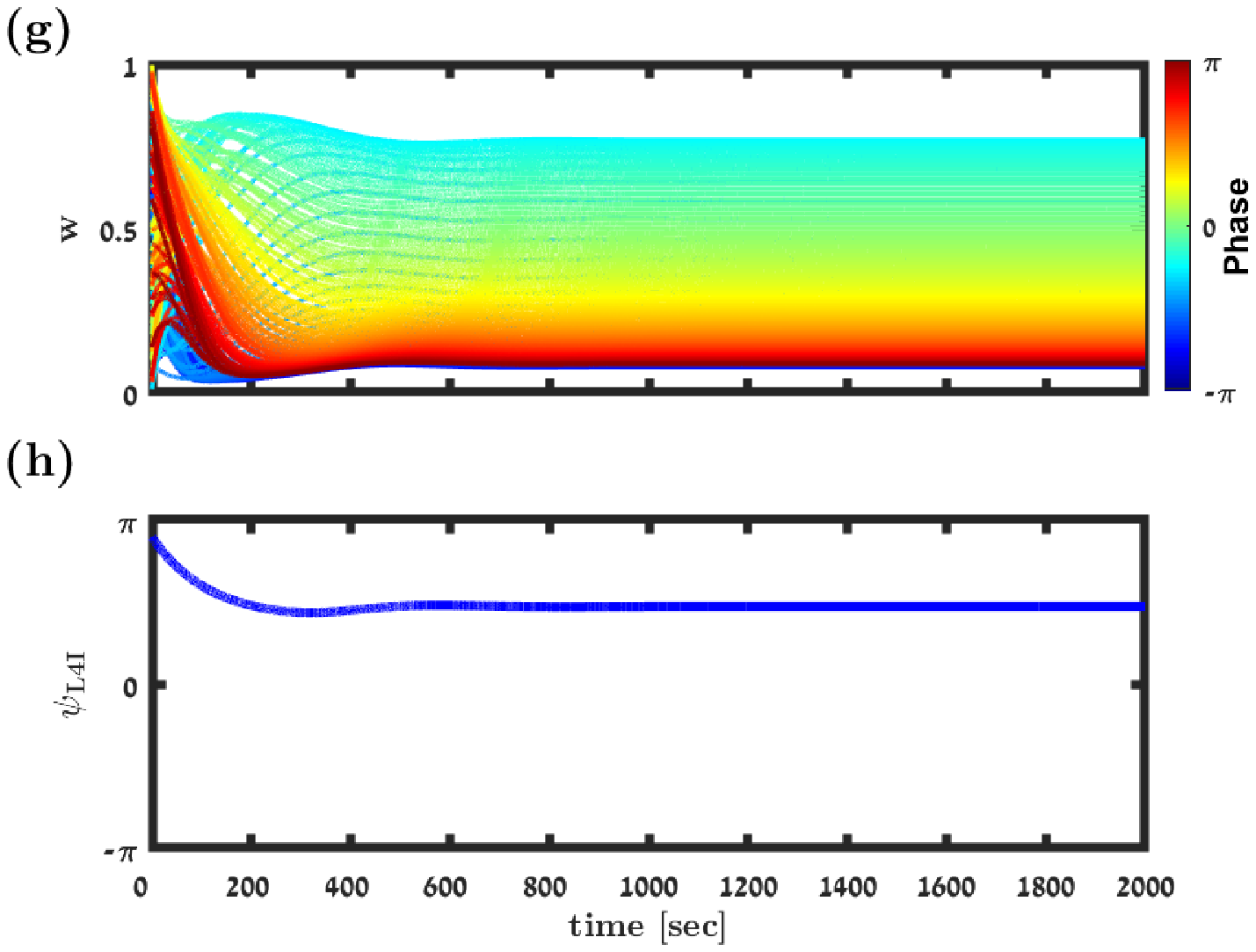} 
			\caption{} 
		\end{subfigure}
		\caption{ \textbf{Transition to a fixed point solution}.  (a), (c), (e) \& (f) show the synaptic weight dynamics for different values of $\mu$.
		The synaptic weights are depicted by different traces, colored according to the preferred phase of the pre-synaptic neuron, see legend. 
		(b), (d), (f) \& (h) show the preferred phase of the downstream L4I neuron, $\psi_{ \mathrm{L4I}}$, as a function of time.
		The parameters used in these simulation were: $N = 150$,  $\kappa=1$, $\bar{\nu}=\nu/(2 \pi)= 7\text{Hz}$, $\gamma=0.9$, $D = 10 \text{Hz}$, and $d=3 \text{ms}$. 
		We used here the temporally asymmetric learning rule, \cref{eq:kernel}, with: $\tau_-=50\text{ms}$, $\tau_+=22\text{ms}$, and $\alpha=1.1$. For the non-linearity parameter we used: $\mu=0.01$ in  (a) \& (b),  $\mu=0.06$. 	 in	(c) \& (d),   $\mu=0.07$ in (e) \& (f). and	  $\mu=0.1$ in (g) \& (h). }
				\label{fig:Fig8}
	\end{adjustwidth} 
\end{figure}

\begin{figure*}[tb!]
			
			\centering

					\begin{subfigure}[t]{0.007\textwidth}
					\textbf{(a)} 
				\end{subfigure}
				\begin{subfigure}[t]{0.42\textwidth}  \vspace{0.007\textwidth}
					\includegraphics[width=\linewidth, valign=t]{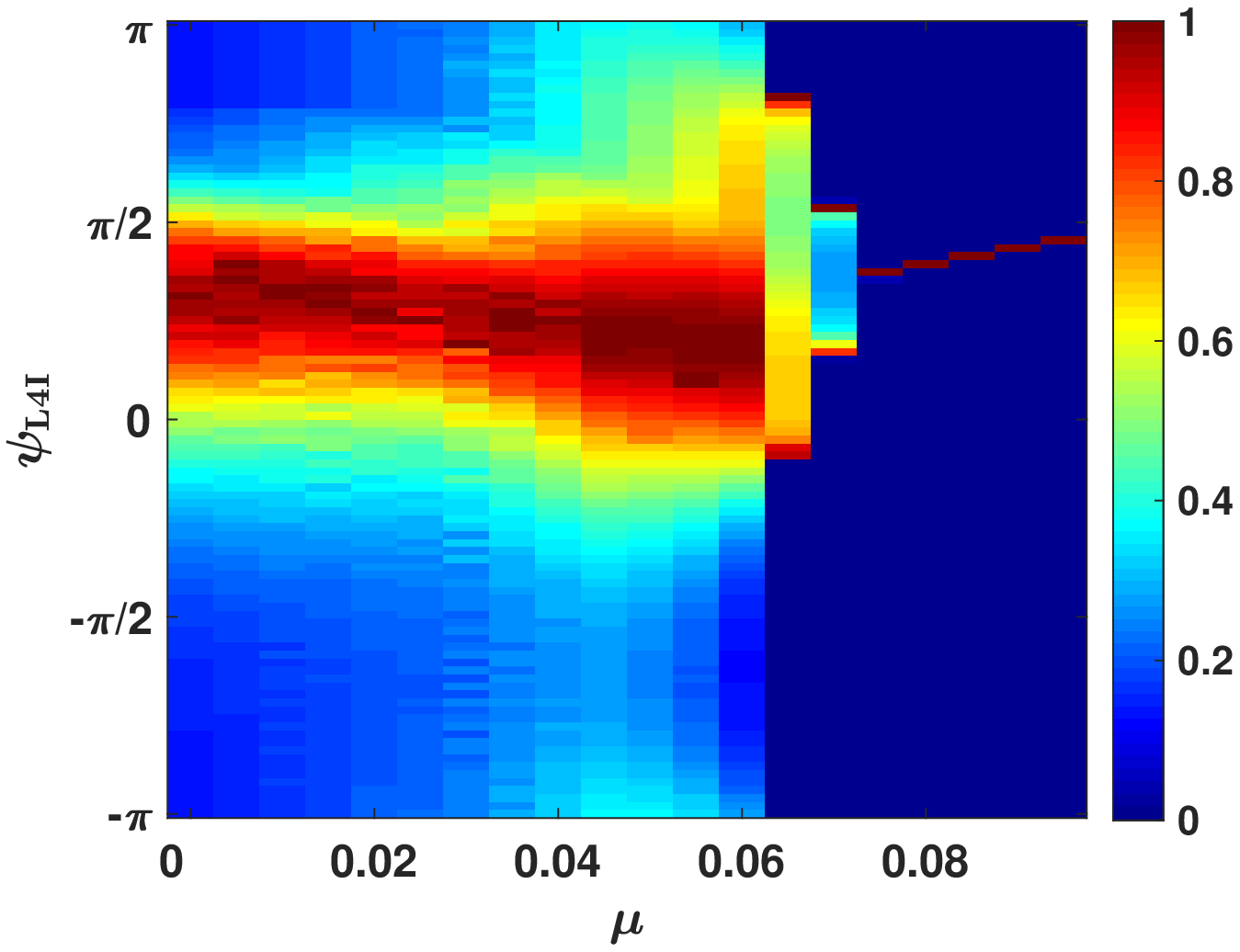}\\[3pt]
					\caption{}
					\label{fig:Fig9a}
				\end{subfigure}\hspace{0.5cm} \vspace{0.02\textwidth}
				\begin{subfigure}[t]{0.007\textwidth}
				\textbf{(b)} 
			\end{subfigure}
			\begin{subfigure}[t]{0.42\textwidth}  \vspace{0.007\textwidth}
				\includegraphics[width=\linewidth, valign=t]{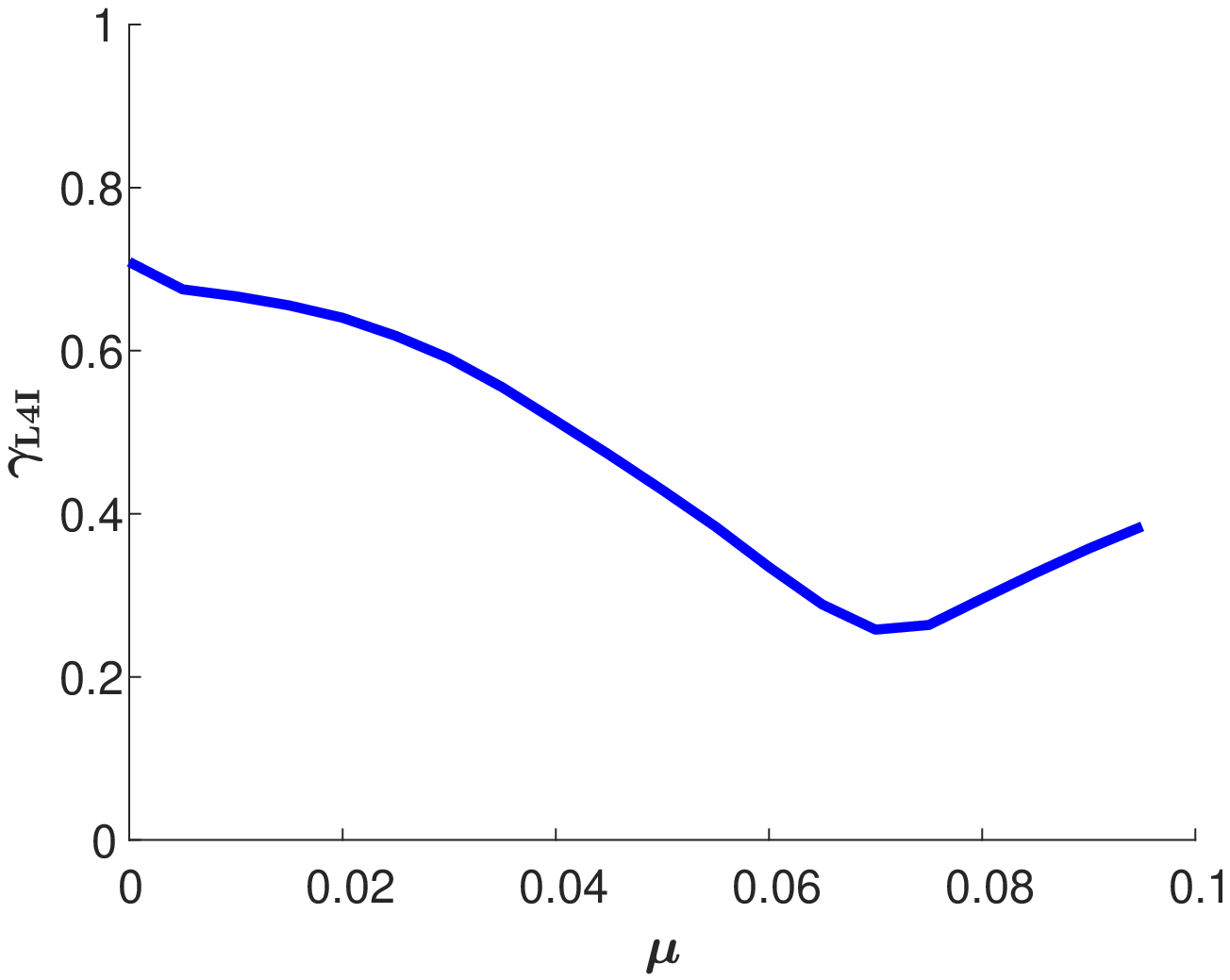}\\[3pt]
				\caption{}
				\label{fig:Fig9b}
			\end{subfigure}\hspace{0.5cm} 
			\hfill \vspace{0.0001\textwidth}
			
			\caption{ \textbf{The effect of the non-linearity parameter, $\mu$.} 
			(a) The distribution of preferred phases of L4I neurons, $\psi_{ \mathrm{L4I}}$, is shown by color as a function of  $\mu$. 
			(b) The modulation depth, $\gamma_{ \mathrm{L4I}}$, is depicted  as a function of $\mu$.  
			The parameters used here were: $\kappa_\mathrm{VPM}=1$, $\psi_{ \mathrm{VPM} }= 5 \pi/6$, $\gamma=0.9$, $D = 10 \text{hz}$, and $d=3 \text{ms}$.
			The temporally asymmetric STDP rule, \cref{eq:kernel}, was used with: $\tau_-=50\text{ms}$, $\tau_+=22\text{ms}$, $\alpha=1.1$ and $\lambda=0.01$.}  
			\label{fig:Fig9}
	\end{figure*}

\subparagraph{The effects of synaptic weight dependence, $\mu$.}
Previous studies have shown that increasing $\mu$ weakens the positive feedback of the STDP dynamics, which generates multi-stability, and stabilizes more homogeneous solutions \cite{gutig2003learning,morrison2008phenomenological,luz2016oscillations,sherf2020multiplexing}. This transition is illustrated in \cref{fig:Fig8}: at low values of $\mu$, the STDP dynamics converges to a limit cycle in which both the synaptic weights and the phase of the the L4I neuron cover their entire dynamic range, \hyperlink{Fig8}{Fig.\ 8a and 8b}. As $\mu$ is increased, the synaptic weights become restricted in the limit cycle and no longer span their entire dynamic range, \hyperlink{Fig8}{Fig.\ 8c and 78}. A further increase of $\mu$ also restricts  the phase of the L4I neuron along the limit cycle, \hyperlink{Fig8}{Fig.\ 8e and 8f}. Finally, when $\mu$ is sufficiently large, STDP dynamics converge to a fixed point, \hyperlink{Fig8}{Fig.\ 8g and 8h}. This fixed point selectively amplifies certain phases, yielding a higher value of $\gamma$ than in the uniform solution. 
These results are summarized in \cref{fig:Fig9} that shows the (non-normalized) distribution of preferred phases and the relative strength of the rhythmic component, $\gamma_{ \mathrm{L4I}}$, as a function of $\mu$. Note that except for a small value range of $\mu$, around which the distribution of preferred phases is bi-modal (see, e.g.\ \hyperlink{Fig8}{Fig.\ 8e and 8f} and $\mu \approx 0.07$ in \cref{fig:Fig9}), STDP yields higher $\gamma_{ \mathrm{L4I} }$ values than in the uniform pooling model.

\begin{figure*}[tb!]
			
			\centering

					\begin{subfigure}[t]{0.007\textwidth}
					\textbf{(a)} 
				\end{subfigure}
				\begin{subfigure}[t]{0.42\textwidth}  \vspace{0.007\textwidth}
					\includegraphics[width=\linewidth, valign=t]{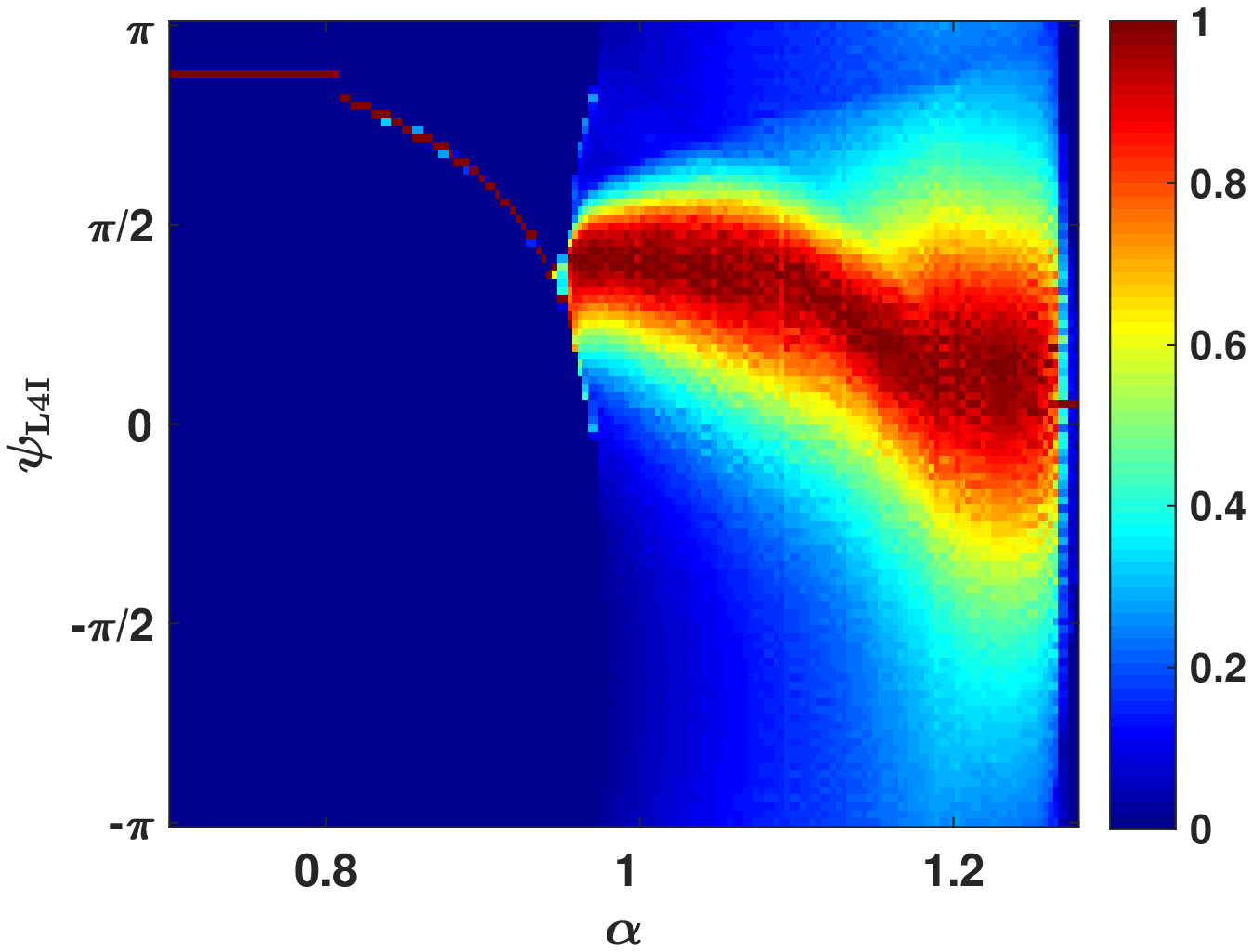}\\[3pt]
					\caption{}
					\label{fig:Fig10a}
				\end{subfigure}\hspace{0.5cm} \vspace{0.02\textwidth}
				\begin{subfigure}[t]{0.007\textwidth}
				\textbf{(b)} 
			\end{subfigure}
			\begin{subfigure}[t]{0.42\textwidth}  \vspace{0.007\textwidth}
				\includegraphics[width=\linewidth, valign=t]{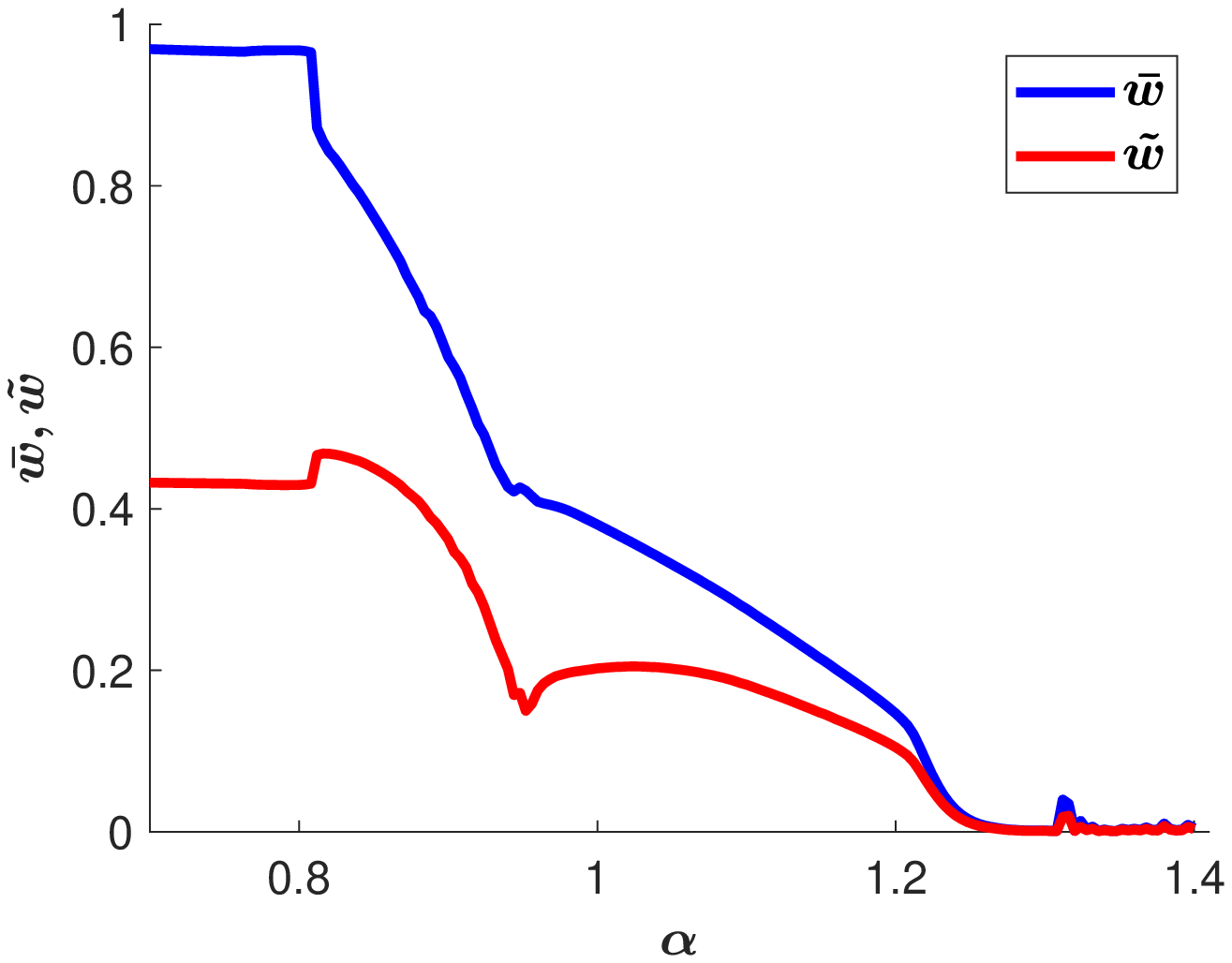}\\[3pt]
				\caption{}
				\label{fig:Fig10b}
			\end{subfigure}\hspace{0.5cm} 
			\hfill \vspace{0.0001\textwidth}

			\caption{ \textbf{The effect of the relative strength of depression, $\alpha$.} 
			(a) The mean input phase $\psi_\mathrm{L4I}$ as a function of  $\alpha$. Probability of occurrences is depicted by color as shown in the right color bar. 
			(b) The order parameters $\bar{w}$ and $\tilde{w}$ are shown as a function of $\alpha$. 
			Here, we used: $\kappa_\mathrm{VPM}=1$, $\psi_0= 5 \pi/6$, $\gamma=1$, $D = 10 \text{hz}$, and $d=3 \text{ms}$. 
			The temporally asymmetric STDP rule, \cref{eq:kernel}, was used with: $\tau_-=50\text{ms}$, $\tau_+=22\text{ms}$, $\mu=0.01$ and $\lambda=0.01$.}  
			\label{fig:Fig10}
	\end{figure*}

\subparagraph{The relative strength of depression, $\alpha$.}
As one might expect, decreasing $\alpha$ beyond a certain value will result in a potentiation dominated STDP dynamics, saturating all synapses at a proximity to their maximal value. Thus, approaching to the uniform solution, which is characterized by a narrow preferred phase distribution centered around the mean preferred phase of the VPM neurons shifted by the delay, and low values of $\gamma_{ \mathrm{L4I} }$, \cref{fig:Fig10}.
Increasing $\alpha$ strengthens the competitive winner-take-all like nature of the STDP dynamics. Initially, this competition will generate a fixed point with non-uniform synaptic weights; thus increasing $\gamma_{ \mathrm{L4I} }$. A further increase of $\alpha$ results in a limit cycle solution to the STDP dynamics that widens the distribution of the preferred phases. Increasing $\alpha$ beyond a certain critical value will result in depression dominated STDP dynamics, driving the synaptic weights to zero,  \cref{fig:Fig10}.

\begin{figure} \hypertarget{fig tau}{}
	\begin{subfigure}[t]{0.007\textwidth} \vspace{-3.8cm}
				\textbf{(a)} 
			\end{subfigure}
		\begin{subfigure}{0.45\columnwidth}
			\includegraphics[width=\textwidth]{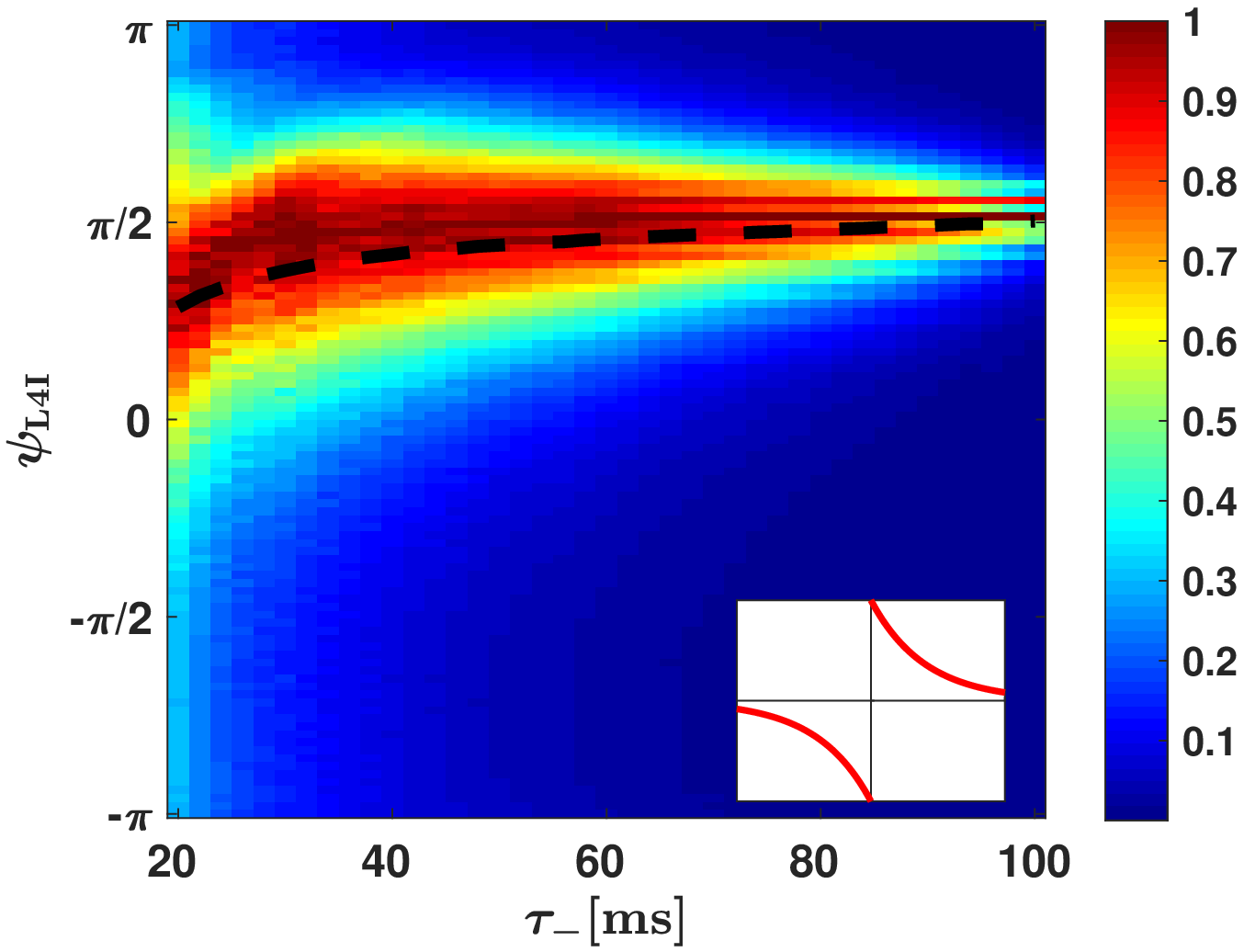}
			\caption{}
		\label{fig:Fig11a}
		\end{subfigure}
		\hfill
			\begin{subfigure}[t]{0.007\textwidth} \vspace{-3.8cm}
			\textbf{(b)} 
		\end{subfigure}
			\begin{subfigure}{0.45\columnwidth}
			\includegraphics[width=\textwidth]{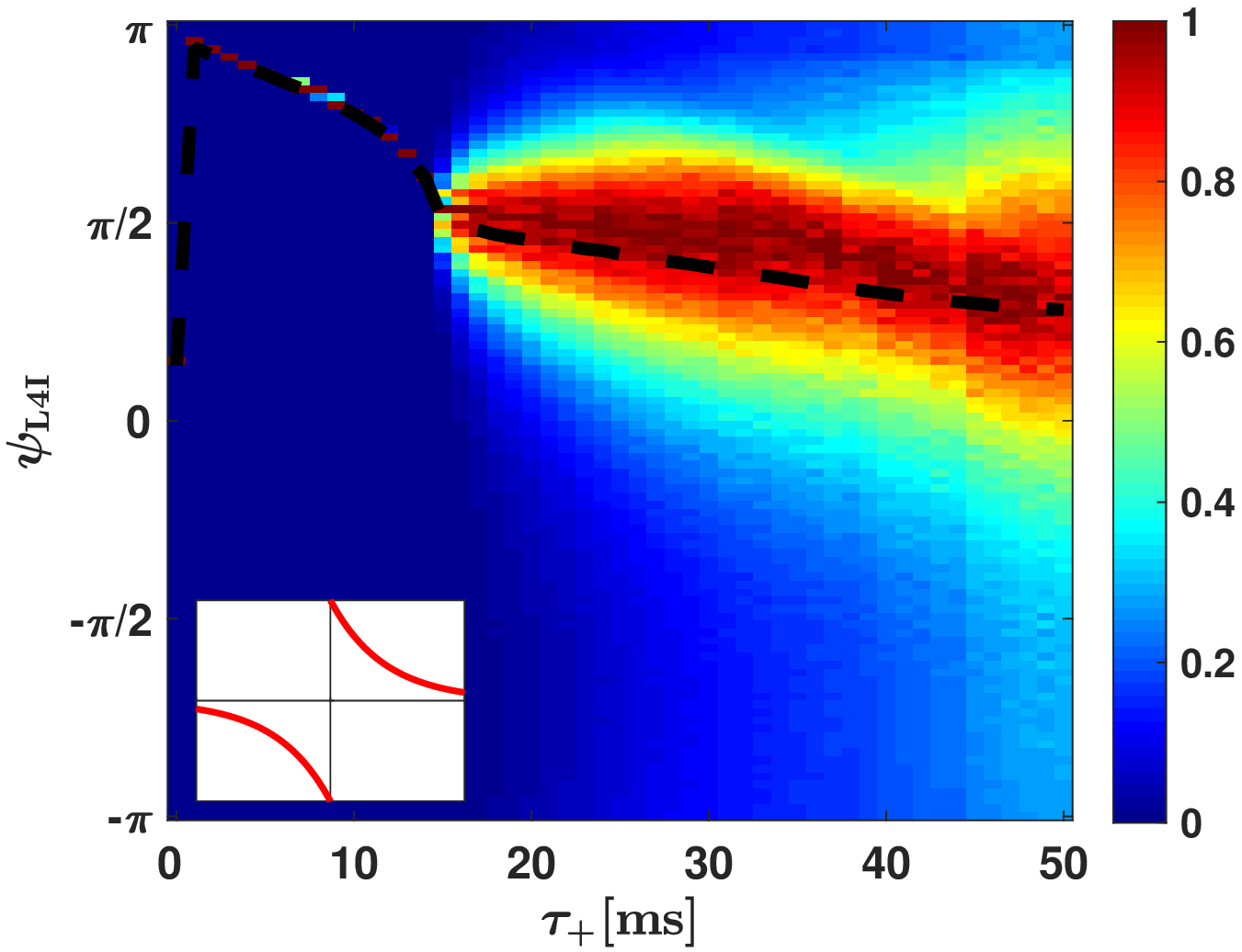}
			\caption{} 
			\label{fig:Fig11b}
		\end{subfigure} 
		
			\begin{subfigure}[t]{0.007\textwidth} \vspace{-3.8cm}
			\textbf{(c)} 
		\end{subfigure}
		\begin{subfigure}{0.45\columnwidth} 
			\includegraphics[width=\textwidth]{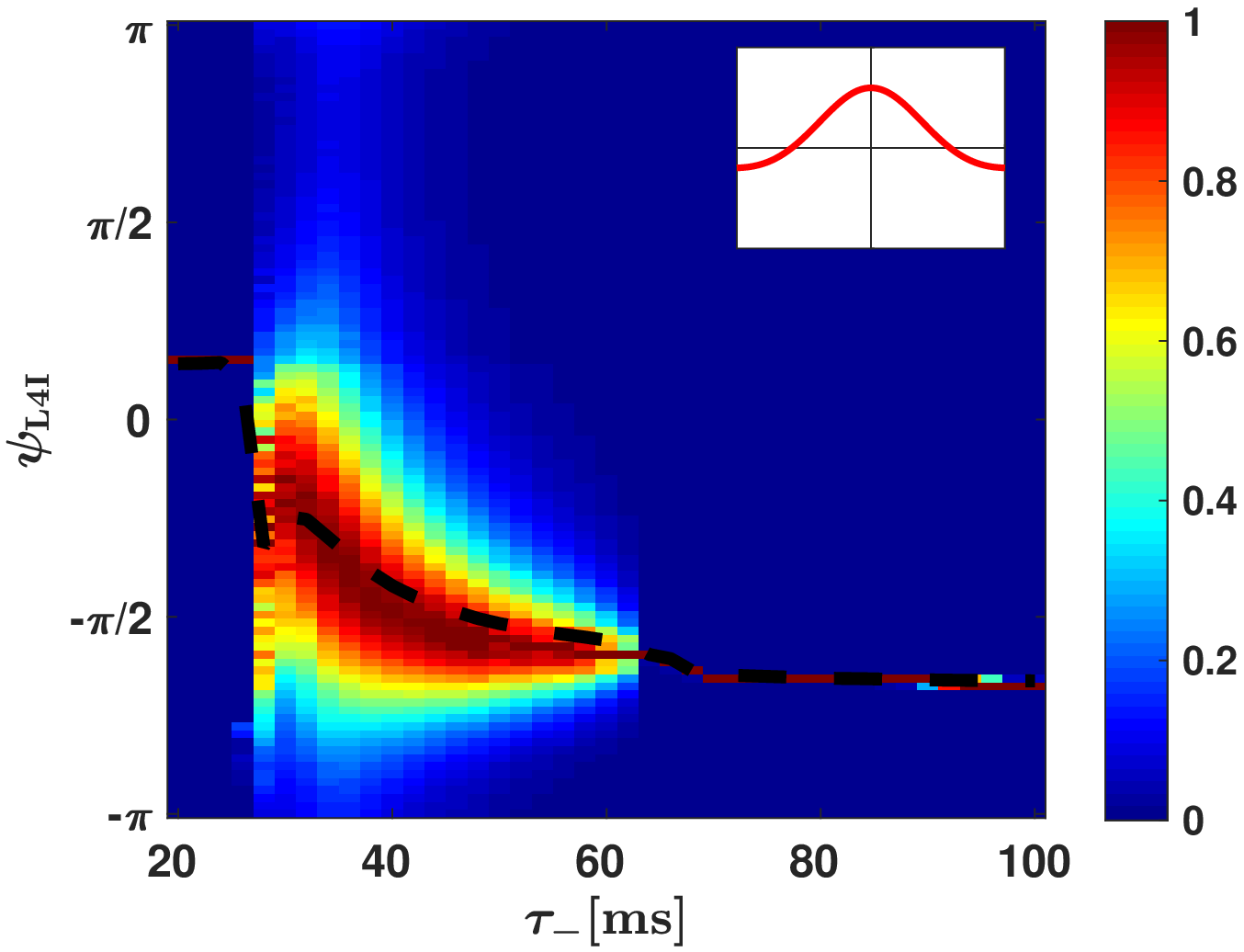} 
			\caption{} 
			\label{fig:Fig11c}
		\end{subfigure}  
		\hfill 
			\begin{subfigure}[t]{0.007\textwidth} \vspace{-3.8cm}
			\textbf{(d)} 
		\end{subfigure}
		\begin{subfigure}{0.45\columnwidth} 
			\includegraphics[width=\textwidth]{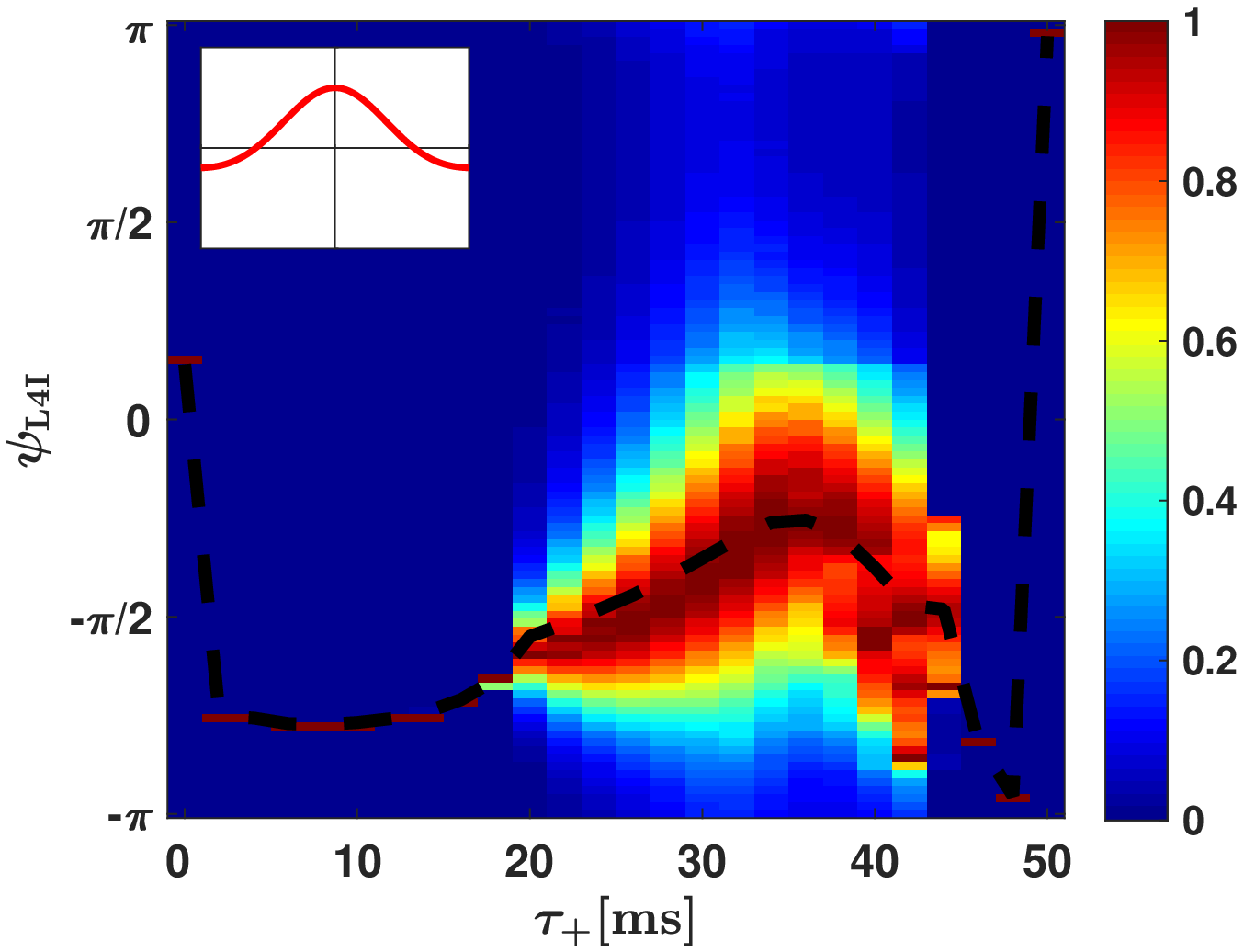} 
			\caption{} 
				\label{fig:Fig11d}
		\end{subfigure}
		\caption{ \textbf{The effect of the temporal structure of the STDP rule.} 
		The distribution of preferred phases of  L4I neurons, $\psi_\mathrm{L4I}$, is shown as a function of the characteristic timescales of the STDP: $\tau_-$ in (a) \& (c) and $\tau_-+$ in (b) \& (d), for the temporally asymmetric rule in (a) \& (b) and the symmetric rule in (c) \& (d). The dashed black lines depict the mean phase. 
		Unless stated otherwise, the parameters used here were: $\kappa_{ \mathrm{VPM}}= 1$, $\gamma=1$, $D = 10 \text{hz}$, $\tau_-=50\text{ms}$, $\tau_+=22\text{ms}$, $d=10 \text{ms}$, $\psi_{ \mathrm{VPM}} = 5 \pi/6$, $\mu=0.01$, $\alpha=1.1$ and $\lambda=0.01$.}
				\label{fig:Fig11}
\end{figure}

\subparagraph{Parameters characterizing the temporal structure of STDP.}
\cref{fig:Fig11} shows the effect of the temporal structure of the STDP on the distribution of preferred phases in the downstream population. As can be seen from the figure, varying the characteristic timescales of potentiation and depression, $\tau_+$ and $\tau_-$, induces quantitative effects on the distribution of the preferred phases. However, qualitative effects  result from changing the nature of the STDP rule. Comparing the temporally symmetric rule, \hyperlink{fig tau}{Fig. 11a and 11b}, with the temporally symmetric rule, \hyperlink{fig tau}{Fig. 11c and 11d}, reveals a dramatic difference in the mean preferred phase of L4I neurons, dashed black lines.

	\section*{Discussion} \label{Conclusion}
We studied the possible contribution of STDP to the diversity of preferred phases of whisking neurons. Whisking neurons can be found along different stations throughout the somatosensory information processing pathway \cite{severson2017active,moore2015vibrissa,wallach2016going,szwed2003encoding,isett2020cortical,ni2017long,gutnisky2017mechanisms,yu2016layer,diamond2008and}. Here we focused on L4I whisking neurons that receive their whisking input mainly via excitatory inputs from the VPM, and suggested that the non-trivial distribution of preferred phases of L4I neurons results from a continuous process of STDP.

STDP has been reported in thalamocortical connections in the barrel system. However, STDP has only been observed during early stages of development \cite{itami2012developmental,itami2016developmental,kimura2019hypothetical}. Is it possible that STDP contributes to shaping thalamocortical connectivity only during the early developmental stages? This is an empirical question that can only be answered experimentally. Nevertheless, several comments should be made. 

First, Inglebert and colleagues recently showed that pairing single pre- and post-synaptic spikes, under normal physiological conditions, does not induce plastic changes \cite{inglebert2020synaptic}. On the other hand, activity dependent plasticity was observed when stronger activity was induced. Thus, it is possible that whisking activity, which is a strong collective and synchronized activity, may induce STDP of thalamocortical connectivity. 

Second, in light of the considerable volatility of synaptic connections observed in the central nervous system \cite{loewenstein2011multiplicative,mongillo2017intrinsic,mongillo2018inhibitory,ziv2018synaptic}, it is hard to imagine that thalamocortical connectivity will remain unchanged throughout the lifetime of the animal. Third, if only activity independent plasticity underlies synaptic volatility, then one expects thalamocortical synaptic weights to be random. In this case, thalamic whisking input to layer 4 should be characterized by an extremely narrow distribution centered around the delayed mean thalamic preferred phase. As layer 4 excitatory neurons have been reported to exhibit very weak rhythmic activity \cite{yu2016layer,gutnisky2017mechanisms}, it is not clear that layer 4 recurrent dynamics can generate a considerably wider distribution of preferred phases that is not centered around the (delayed) mean phase in the VPM.

Consequently, a non-trivial distribution of L4I phases, with $\kappa_{ \mathrm{L4I}} \sim 1$, can be obtained either via STDP or by pooling the whisking signal from an extremely small VPM population of $N < 10$ neurons. However, in the latter scenario the mean preferred phase of L4I neurons is expected to be determined by the (delayed) mean phase of VPM neurons, thus raising serious doubts as to the viability of the latter solution.

Our hypothesis views STDP as a continuous process. Functionality, in terms of transmission of the whisking signal and retaining stationary distribution of the preferred phases in the downstream population, is obtained as a result of continuous remodelling of the entire population of  synaptic weights. The distribution of preferred phases of L4I neurons reflects the distribution of the preferred phase of a single neuron over time. This key feature of our hypothesis provides a clear empirical prediction. By monitoring the preferred phases of single L4I neurons, our theory predicts that these phases will fluctuate in time with a non-uniform drift velocity that depends on the phase. Our theory predicts a direct link between this drift velocity and the distribution of preferred phases of L4I neurons. Additionally, our theory draws a direct link between the STDP rule and the distribution of preferred phases in the downstream population (see e.g.\ \cref{fig:Fig11}), which, in turn, serves as further prediction of our theory. 

In the current study, we made several simplifying assumptions. The spiking activity of the thalamic population was modeled as a rhythmic activity with a well-defined frequency. However, whisking activity spans a frequency band of several Hertz. Moreover, the thalamic input relays additional signals, such as touch and texture. These signals will modify the cross-correlation structure and will add `noise' to the dynamics of the preferred phase of the downstream neuron. As a result, the distribution of preferred phases in the downstream population is expected to widen. In addition, our analysis used a purely feed-forward architecture ignoring recurrent connections in layer 4, which may also affect the preferred phases in layer 4. A quantitative application of our theory to the whisker system should consider all these effects. Nevertheless, the theory presented provides the basic foundation to address these effects.

	\section*{Methods} \label{Methods}

		\subsection*{Temporal correlations} \label{subsec:correl}
	
	 The cross-correlation between pre-synaptic neurons $j$ and $k$ at time difference $\Delta t$ is given by:
	\begin{equation}\label{eq:correprepre}
	\begin{split}
	\Gamma_{(j, k)}(\Delta t)&=\langle  \rho_{j}(t) \rho_{k}(t+\Delta t)   \rangle=  D^2  \big(1  + 
	\frac{\gamma^2}{2} \cos[ \nu \Delta t\\&
	+\phi_{j}-\phi_{k}]\big)+ \delta_{j k} D \delta(\Delta t).
	\end{split}
	\end{equation}

	In the linear Poisson model, \cref{eq:meanpost}, the cross-correlation between a pre-synaptic neuron and the post-synaptic neuron can be written as a linear combination of the cross-correlations in the upstream population; hence,  
	the cross-correlation between the $j$th VPM neuron and the post-synaptic neuron is
	\begin{equation}\label{eq:corrFull}
	\begin{split}
	\Gamma_{j, \text{ post}}(\Delta t) &= \frac{D}{N}\delta(\Delta t-d) w_{j} +
	D^2 
	\bigg( \bar{w}  + 
	\frac{\gamma^2}{2} \tilde{w} \cos[ \nu (\Delta t-d) \\ &+ \phi_{j}-  \psi ]\bigg).
	\end{split}
	\end{equation}
   Where  $\bar{w}$ and  $\tilde{w}e^{i \psi}$ are order parameters characterizing the synaptic weights profile, as defined in \cref{eq:wbar,eq:wtilda}.

		\subsection*{The mean field Fokker-Planck dynamics} \label{Approximated model}

		For large $N$ we obtain the continuum limit from  \cref{eq:wdot}:

		\begin{equation}\label{eq:wdotCont}
		\frac{\dot{w}(\phi,t)}{\lambda}=F_{d}(\phi,t)+\bar{w}(t)F_{0}(\phi,t)+\tilde{w}(t)F_{1}(\phi,t),
		\end{equation}
		where 
		\begin{subequations} \label{eq:Fd01}
			\begin{align}
			\begin{split}
			F_{d}(\phi,t)=&w(\phi,t)\frac{D}{N}\bigg(f_{+}(w(\phi,t))K_+(d)-\\
			& f_{-}(w(\phi,t)) K_-(d)\bigg), 
			\end{split}\\      
			F_{0}(\phi,t)=&D^2 \bigg(\bar{K}_{+}f_{+}(w(\phi,t))   -\bar{K}_{-}f_{-}(w(\phi,t))\bigg),\\
			\begin{split}
			F_{1}(\phi,t)=& D^2 \frac{\gamma^2}{2}\bigg(\tilde{K}_+ f_{+}(w(\phi,t)) \cos[\phi-\Omega_+- \\
			& \nu d-\psi]- \tilde{K}_- f_{-}(w(\phi,t)) \cos[\phi-\Omega_-
			\\ &- \nu d-
			\psi]\bigg),
			\end{split}        
			\end{align}
		\end{subequations}
		and $\bar{K}_{\pm}$, $\tilde{K}_{\pm} e^{i \Omega^\eta_{\pm}}$ are the Fourier transforms of the STDP kernels
		\begin{align}\label{eq:Kfourier}
		&\bar{K}_{\pm}=\int_{-\infty}^{\infty}K_{\pm}(\Delta) d\Delta,\\
		&\tilde{K}_{\pm}  e^{i \Omega_{\pm}}=\int_{-\infty}^{\infty}K_{\pm}(\Delta) e^{-i  \nu \Delta} d\Delta.
		\end{align}
		Note that in our specific choice of kernels, $\bar{K}_{\pm}=1$, by construction.

	\subsection*{Fixed points of the mean field dynamics} \label{subsec:Fixed}
	The fixed point solution of \cref{eq:wdotCont} is given by 
	
	\begin{equation}\label{eq:fp}
	w(\phi)^*=\bigg(1+\alpha^{1/\mu}\big(\frac{1+X_-}{1+X_+}\big)^{1/\mu}\bigg)^{-1}, 
	\end{equation}
	where 
	\begin{equation}\label{eq:Xpm}
	X_\pm \equiv	\frac{\tilde{w}}{\bar{w}}\frac{\gamma^2}{2}\tilde{K}_\pm\cos(\phi-\nu d-\Omega_{\pm}-\psi).
	\end{equation}
	Note that, from \Cref{eq:fp} and \cref{eq:Xpm}, the fixed point solution, $w(\phi)^*$, will depend on $\phi$, for $\kappa_{ \mathrm{VPM}} > 0$. As $\mu$ grows to 1 the fixed point solution will become more uniform, see \cite{luz2016oscillations}.

	\subsection*{Details of the numerical simulations \& statistical analysis} \label{Details of the numeric simulations}
	
Scripts of the numerical simulations were written in Matlab. The numerical results presented in this paper were obtained by solving \cref{eq:wdotCont} with the Euler method with a time step $\Delta t=0.1$ and $\lambda=0.01$.
	The cftool with the Nonlinear Least Squares method and a confidence level of $95 \%$ for all parameters was used for fitting the data presented in  \cref{fig:Fig3,fig:Fig5b,fig:Fig6b,fig:Fig7b}.

		\subsection*{Modeling pre-synaptic phase distributions} \label{Details of the phase distribution}

Unless stated otherwise, STDP dynamics in the mean field limit was simulated without quenched disorder. To this end, the preferred phase, $\phi_k$, of the $k$th neuron in a population of $N$ pre-synaptic VPM neurons was set by the condition $ \int_{- \pi} ^{ \phi_k} \Pr ( \varphi) d  \varphi = k/N $.
In \cref{fig:Fig4b,fig:Fig4c} we used the accept-reject method  \cite{chib1995understanding,robert2013monte} to sample the phases.


	\section*{Acknowledgments}
	This research was supported by the Israel Science Foundation (grant No. 300/16).

	%
	%
	%
	
	%
	\bibliography{MyBib}
	%
	%

	


\end{document}